\documentclass[12pt,a4paper,twoside,titlepage]{article}
\usepackage[latin2]{inputenc}
\usepackage[OT4]{fontenc}
\usepackage[dvips]{hyperref} 
\usepackage{url}
\usepackage{latexsym}
\usepackage{amsmath}
\usepackage{amssymb}
\usepackage{amsfonts}
\usepackage{wasysym}
\usepackage{fancyhdr}

\hypersetup{%
pdftitle={Assembly and measurements of the Electromagnetic Calorimeter components for WASA at COSY setup},
pdfsubject={Nuclear Physics},
pdfauthor={Benedykt Jany -- Jagiellonian University},
pdfkeywords={WASA at COSY, IKP, COSY, SEC, Calorimeter},
bookmarksnumbered,
pdfstartview={FitV},
}%

\usepackage{fancybox}
\usepackage[figuresright]{rotating}
\usepackage{graphicx}
\usepackage[german,polish,english]{babel}
\usepackage{longtable}

\graphicspath{{images/}}
\newcommand{\emptydoublepage}{\clearpage\thispagestyle{empty}\cleardoublepage}
\newlength{\fullwidth} 
\newcommand{\myFrameBigFigure}[4][image]
{
\begin{figure}[p]
	\frame{%
		\includegraphics[width=\textwidth]{#2}%
		}
		\caption[#4]{#3}
		\label{#1_#2}
\end{figure}
}

\newcommand{\myFrameHugeFigure}[4][image]
{
\begin{sidewaysfigure}[t!bp]
\frame{\includegraphics[width=20cm, height=13cm]{#2}}		
\caption[#4]{#3}
		\label{#1_#2}
\end{sidewaysfigure}
}
\newcommand{\myFigure}[4][image]%
{%
\begin{figure}[ht!bp]%
	\begin{center}%
		\includegraphics[width= \textwidth]{#2}%
		\caption[#4]{#3}
		\label{#1_#2}%
	\end{center}%
\end{figure}%
}%

\newcommand{\myFrameFigure}[4][image]%
{
\begin{figure}[ht!bp]
	\begin{center}
	\frame{\includegraphics[width= \textwidth]{#2}}
		\caption[#4]{#3}
		\label{#1_#2}
	\end{center}
\end{figure}
}

\newcommand{\myFrameSmallFigure}[4][image]%
{
\begin{figure}[ht!bp]
	\begin{center}
	\frame{\includegraphics[width=0.7\textwidth]{#2}}
		\caption[#4]{#3}
		\label{#1_#2}
	\end{center}
\end{figure}
}
\newcommand{\myFrameSmallFigurer}[4][image]%
{%
\begin{figure}[ht!bp]
	\begin{center}
	\frame{\includegraphics[height= 0.7\textwidth, angle=270]{#2}}
		\caption[#4]{#3}
		\label{#1_#2}
	\end{center}
\end{figure}
}%

\newcommand{\myImgRef}[2][image]%
{%
	Fig.~\ref{#1_#2}%
}
\newcommand{\myTable}[3]%
{%
\begin{table}[htdp]%
	\begin{center}%
		#1%
		\caption{#2}%
		\label{#3}%
	\end{center}%
\end{table}%
}%

\newcommand{\myTabRef}[1]%
{%
	Tab.~\ref{#1}%
}%
\newcommand{\myEqRef}[1]%
{%
	Eq.~\ref{#1}%
}%
\begin{document}

\selectlanguage{english}

\begin{titlepage}

\begin{center}
\begin{Huge}Jagiellonian University\\
\end{Huge}
\begin{center}
\textsc{Marian Smoluchowski Institute of Physics}
\end{center}
\end{center}
\vspace{0.5cm}
\begin{center}
\includegraphics[width=4.5cm]{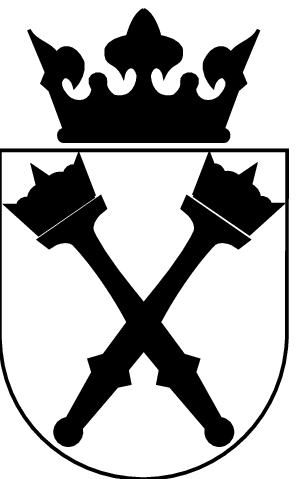}
\end{center}
\vspace{1.5cm}


\begin{center}
\begin{Large}
{Assembly and measurements} \\ {of the Electromagnetic Calorimeter components} \\ {for ``WASA~at~COSY'' setup}
\end{Large}
\end{center}
\vspace{0.5cm}
\begin{center}
\textbf{\begin{Large}Benedykt Jany\end{Large}}
\end{center}

\vspace{1.5cm}
\begin{flushright}
Master Thesis prepared at the Nuclear Physics Department\\
Supervisor: Dr. hab. Zbigniew Rudy
\end{flushright}
\vspace{2cm}

\begin{center}
\textbf{\begin{large}Cracow $2006$\end{large}}
\end{center}

\emptydoublepage
\thispagestyle{empty}
\begin{tiny}$\infty$\end{tiny}\\
\vspace{0.5\textheight}

\begin{Large}\textsc{``Quod scimus, gutta est, ignoramus mare.''}\end{Large}

\end{titlepage}

\emptydoublepage

\begin{abstract}
This work describes the tests of the scintillator electromagnetic calorimeter of the WASA detector setup after transferring it from the CELSIUS storage ring at The Svedberg Laboratory(Uppsala, Sweden) to the Cooler Synchrotron COSY at the Institut f\"ur Kernphysik (IKP) of Forschungszentrum J\"ulich, Germany. The tests were performed using gammas of $4.4$MeV energy from an AmBe source. The status of the CsI(Na) crystals was determined and the indication of gain factor for the energy calibration was extracted. 
\end{abstract}

\emptydoublepage
\begin{otherlanguage}{polish}
  \begin{abstract}
Ta praca magisterska opisuje testy kalorymetru elektromagnetycznego detektora WASA po jego transferze z akceleratora CELSIUS The Svedberg Laboratory(Uppsala, Szwecja) do Forschungszentrum J\"ulich Institut f\"ur Kernphysik (IKP) Niemcy (akcelerator COSY). Testy zostały przeprowadzone przy użyciu źródła Amerykowo-Berylowego emitującego promienie gamma o energii $4.4\,\mathrm{MeV}$. Kryształy CsI(Na) zostaly sprawdzone oraz stałe kalibracyjne zostały wyliczone.
  \end{abstract}
\end{otherlanguage}

\emptydoublepage

\tableofcontents
\thispagestyle{empty}
\emptydoublepage

\section{Introduction}
This work was performed during my stay from November 2005 to April 2006 in the Institut f\"ur Kernphysik (IKP) of Forschungszentrum J\"ulich, Germany as a diploma student of the Jagiellonian University Cracow. It was motivated by the transfer of the WASA\cite{ekstrom} detector setup from the CELSIUS storage ring at TSL(Uppsala, Sweden) to the Cooler Synchrotron COSY at IKP \cite{cern}. 
This latter results in an unique scientific opportunities for hadron physics with hadronic probes: COSY with its phase-space cooled proton and deuteron  beams covering an energy region up to the $\phi$-meson sector is combined with WASA, a close to $4\pi$ detector for photons and charged particles.
After the transfer to J\"ulich and prior to installation in the COSY ring - which is scheduled for summer 2006 - all detectors needed to be tested and to be tuned. This thesis deals with the electromagnetic calorimeter of WASA, probably the most important and most crucial part of the whole setup.

The thesis is divided into three main parts. Section~\ref{sec:project} describes the advantages of the transfer WASA to J\"ulich\cite{proposal}, and gives an overview over the WASA detector including the pellet target system. Special attention was put to the central part and particularly to the electromagnetic calorimeter.

Section~\ref{sec:calorimetry} reveals the physics of electromagnetic interactions of photons and electrons with matter. Basic issues of calorimeter physics are given. 

Section~\ref{sec:data} describes the performed measurements on the calorimeter. Section~\ref{subsec:electronics} shows how the data were taken and a description of electronics is presented. The general analysis of the data is introduced in section~\ref{subsec:generalAnl}.
Finally section~\ref{subsec:sourceMeas} is dealing with the main goal of this thesis, the test of the CsI(Na) calorimeter crystals with $4.4$MeV gamma source. Section~\ref{subsec:cosmics} provides a comparison with energy deposits by cosmic muons in the calorimeter medium, using a Monte-Carlo simulation of the full WASA setup.

A Summary with conclusions are given in section~\ref{sec:summary} -- future implications for the WASA calorimeter are considered.


\emptydoublepage
\section{The~WASA~at~COSY Project \label{sec:project}}
The WASA detector setup had been operated since 1998. It has been built with the focus on studying the decays of $\eta$-mesons in nuclear reactions. However, in 2003 it has been decided to shut down the CELSIUS ring thus, to stop the operation of WASA~at~CELSIUS. Shortly after that announcement the idea came up to continue the operation of WASA~at~COSY. The reasons are obvious: The combination of WASA and COSY would be of advantages for both communities. COSY offers a higher energy than CELSIUS, allowing the extension of the studies into the $\eta'$ sector. The WASA detector had an electromagnetic calorimeter as a central component and, thus, the ability to detect neutral decay modes involving photons -- such a device is missing at COSY. The proposal \cite{proposal} for moving WASA to COSY was accepted by the COSY PAC in 2004. The WASA detector was dismounted during summer 2005 and shipped to J\"ulich. The final installation is scheduled for summer 2006. 
\subsection{Cooler Synchrotron COSY}
\label{subsec:cosy}
\myFrameSmallFigure{ikpcosy}{View at the Cooler Synchrotron COSY.}{View at the  Cooler Synchrotron COSY}
\newpage
The Cooler Synchrotron COSY \myImgRef{ikpcosy} is located in the IKP of the Forschungszentrum J\"ulich Germany. It delivers phase-space cooled polarized or unpolarized protons (deuterons) of momentum from $p=300$~MeV/c up to $p=3700$~MeV/c. The ring has a circumference of $184$m and can be filled with up to $10^{11}$ particles. When using the internal cluster target the luminosity of value $10^{31}\mathrm{cm^{-2}s^{-1}}$ can be reached. Two cooling methods are applied during accumulation of the beam to reduce the phase-space volume, electron cooling at injection energies and stochastic cooling. In the electron cooling method an electron beam, moving with the same average velocity as proton beam, absorbs the kinetic energy of protons, whereas stochastic cooling is the process in which the deviations of nominal energy or position of particles in a beam are measured and corrected. The electron cooling system in COSY is applied at injection momentum $p=300$~MeV/c and reaches up to $p=600$~MeV/c, and the stochastic cooling from $p=1500$~MeV/c to $3700$~MeV/c.
Both, proton and deuteron beams, can be provided unpolarized as well as polarized. At COSY internal and external target positions are in operation \myImgRef{cosy}. For further details see \cite{cosy}.

\myFrameBigFigure{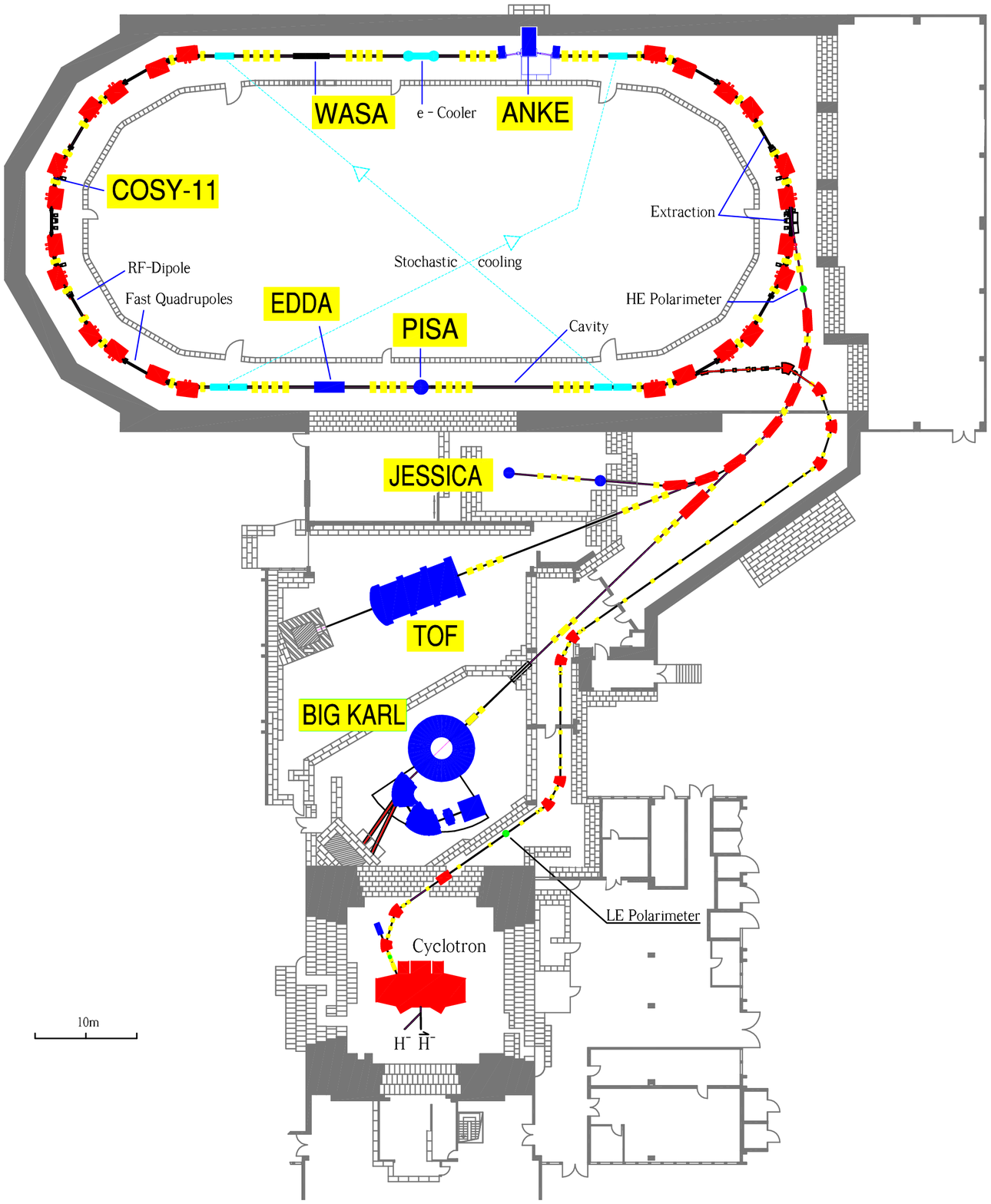}{The accelerator complex with the cyclotron, the COSY ring and the experimental installations. The place of WASA is within one of the straight sections of COSY. For more details please visit \url{http://www.fz-juelich.de/ikp/cosy}.}{The accelerator complex COSY}

\subsection{The WASA detector}
\label{subsec:detector}
\myFrameFigure{wdetector1}{Cross section of the WASA detector as being used at CELSIUS. Beam comes from the left. The Central Detector (\textbf{CD}) build around the interaction point (at the left). The layers of the Forward Detector (\textbf{FD}) are shown on the right. Other symbols (MDC, PSB, FPC, ...) will be explained in text of the thesis.}{Cross section of the WASA detector}
\myFrameSmallFigure{wdetector2}{3D View of the WASA detector\cite{proposal}.}{3D View at the WASA detector}

As mentioned before the \textbf{WASA} detector -- \textit{\textbf{W}ide \textbf{A}ngle \textbf{S}hower \textbf{A}pparatus }  \myImgRef{wdetector1}, was designed to study various decay modes of the $\eta$-meson. This is reflected in the detector setup.
The $\eta$-mesons are produced in reactions of the type $pp \longrightarrow pp\eta$. Due to the kinematics boost, the two protons are going into the forward direction, while the light decay products of the $\eta$ are distributed into $4\pi$.

In order to detect the protons, a $\phi$-symmetric forward detector for $\theta \leq 18^{\circ}$ is installed. The particles are identified and reconstructed by means of dE measurement and track reconstruction using drift chambers. A trigger, which is set only on the forward detector, can be used to select events independently on the decay mode of the $\eta$-meson.

Particles coming from meson decays are $e^{\pm}, \mu^{\pm}, \pi^{\pm}$ and $\gamma$, are detected in the central part of WASA ($\theta \geq 20^{\circ}$). Momentum reconstruction is done by tracking in a magnetic field and the energy of the particles is measured using an electromagnetic calorimeter. By including the central detector in the trigger also very rare decay modes can be studied while using a very high luminosity of up to $10^{32}$~cm$^{-2}$s$^{-1}$. A 3D view of the detector setup is presented in \myImgRef{wdetector2}.

\subsubsection{The Pellet Target}
\label{subsubsec:pellet}
The pellet target system \myImgRef{pellet1} was a special development for WASA. 
The "pellets" are frozen droplets of hydrogen or deuterium with a diameter between $25\mu$m and $35\mu$m. The advantages of using pellet target compared with a standard internal gas target are the following:

\begin{itemize}
\item high target density, allows high luminosities necessary for studying rare decays
\item thin tube delivery through the detector, $4\pi$ detection possible
\item very good localized target, small probability of secondary interactions inside the target
\end{itemize}

The central part of the system is the pellet generator where a stream of liquid gas (hydrogen or deuterium) is broken into droplets by a vibrating nozzle. The droplets freeze by evaporation into a first vacuum chamber forming a pellet beam. The beam enters a vacuum-injection capillary where it is collimated and is fed through a $2\mathrm{m}$ long pipe into the scattering chamber \myImgRef{pellet2}. An effective beam thickness for hydrogen of $3*10^{15}\, \mathrm{atoms/cm^2}$ has been achieved with a beam diameter $2-4\mathrm{mm}$, a frequency of pellets $5-10\mathrm{kHz}$, and an average distance between the pellets of $9-20$mm.

\begin{figure}[ht!bp]%
	\begin{center}%
	\frame{
		\includegraphics[height= 0.7\textwidth, angle=270]{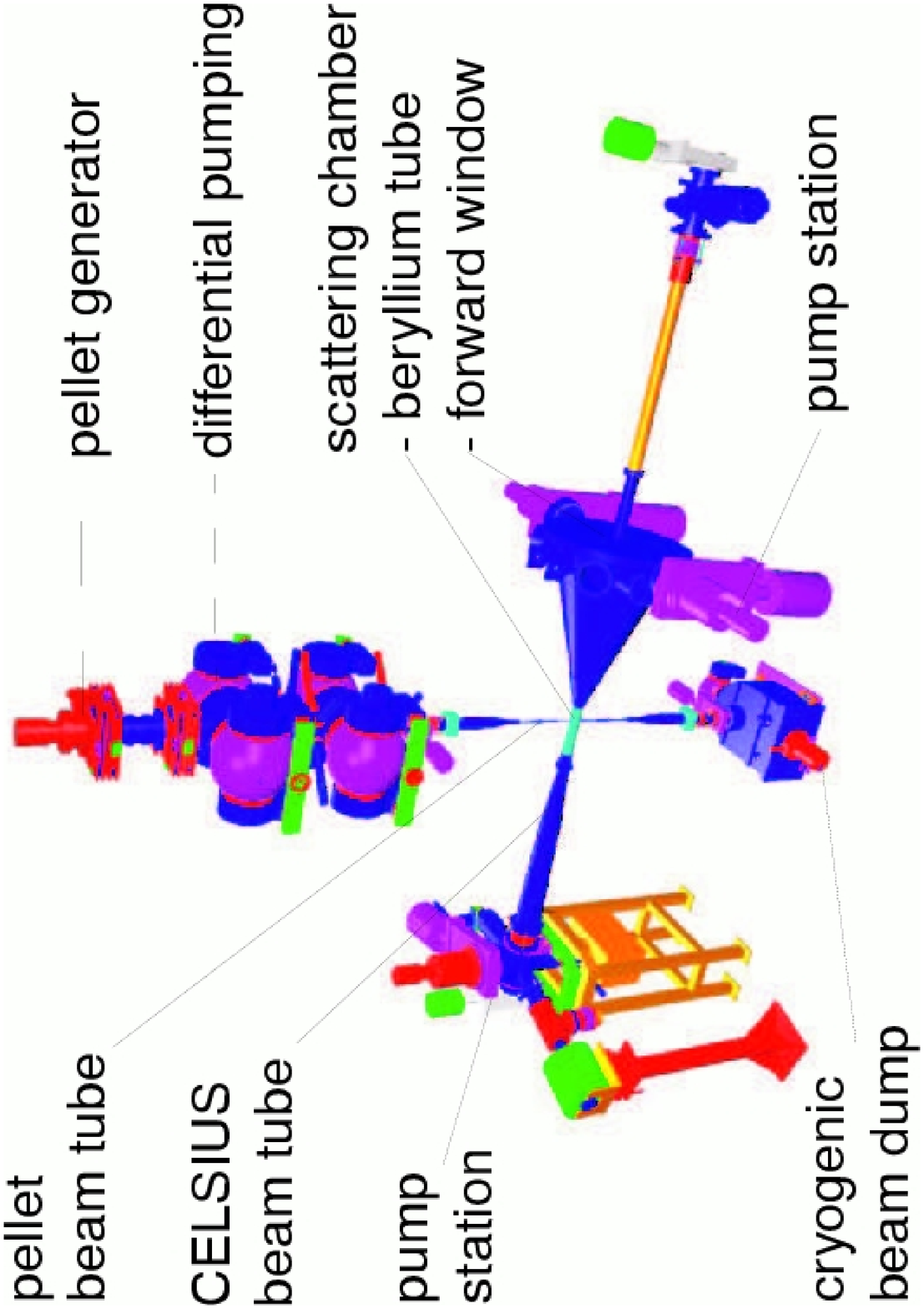}%
		}
		\caption[The Pellet Target system]{The Pellet Target system \cite{proposal}.}
		\label{image_pellet1}%
	\end{center}
\end{figure}
\begin{figure}[ht!bp]
	\begin{center}
	\frame{\includegraphics[width=0.5\textwidth]{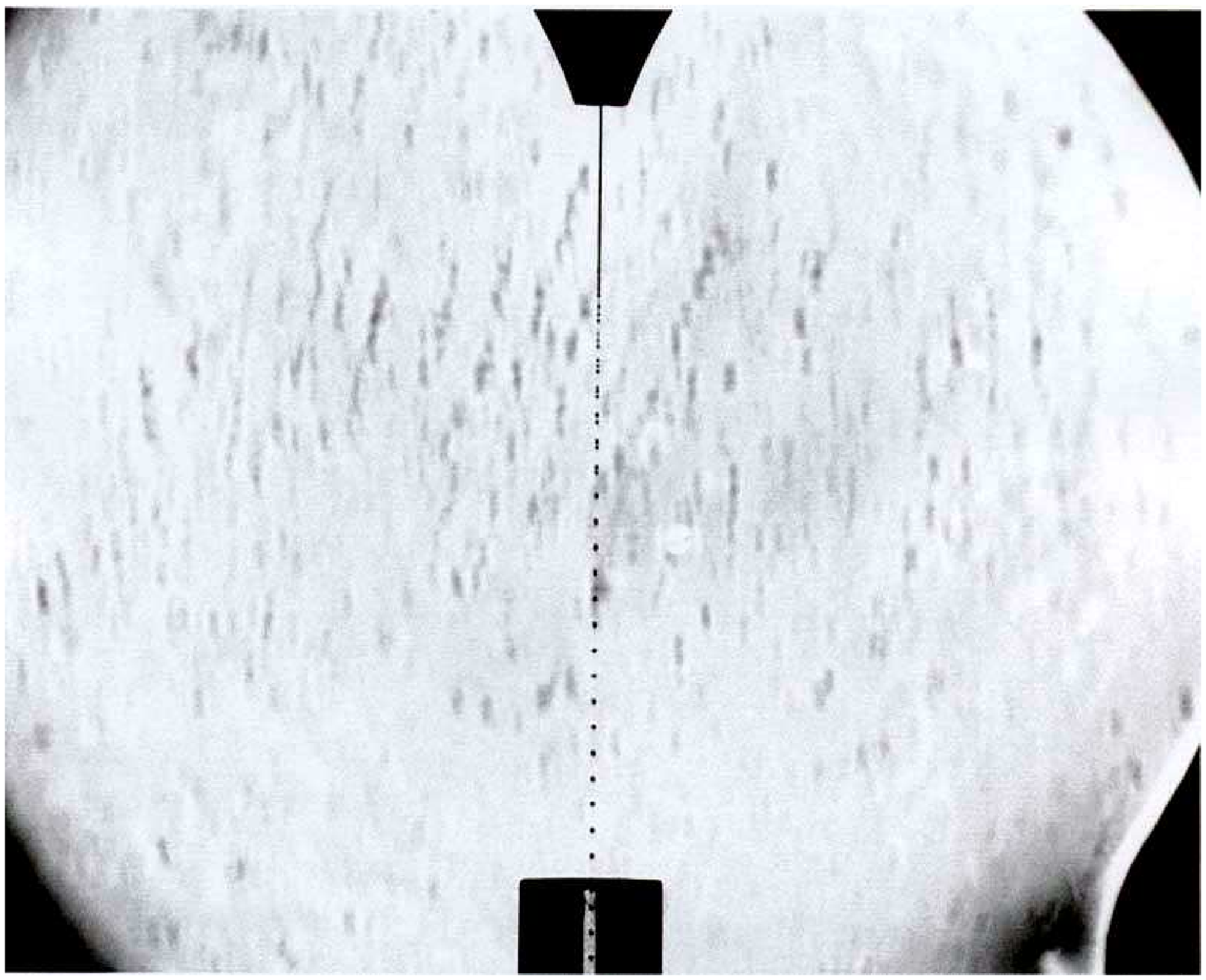}}
		\caption[Photo of generated pellets]{Photo of generated pellets \cite{wwwWASA}.}
		\label{image_pellet2}
	\end{center}
\end{figure}

\newpage
\subsubsection{The Forward Detector}
\label{subsubsec:fd}
The Forward Detector (FD) \myImgRef{wdetector1} tags meson production by measuring the energies (dE-E) and angles of forward scattered projectiles and charged recoil particles like protons and deuterons, also neutrons and charged pions.
The produced mesons are then reconstructed using the missing mass technique. It covers angles from $3^{\circ}$ to $18^{\circ}$. It consists of several layers of detectors described below: 
\begin{itemize}
\item \textbf{FWC} The Forward Window Counters\\
The FWC is the first detector in the Forward Detector. It consists of $12$ plastic scintillators of $5\mathrm{mm}$ thickness. It is used to reduce the background from scattered particles originally from the beam pipe or the exit flange.
\item \textbf{FPC} The Forward Proportional Chambers\\
The next detector is the FPC. It consists of $4$ modules each containing $122$ straw tubes detectors. The modules are rotated relatively to each other by $45^{\circ}$. The FPC is used as a precise tracking device.
\item \textbf{FTH} The Forward Trigger Hodoscope\\
Close to FPC the FTH (``J\"ulich Quirl'') is installed. It consists of 3 layers of plastic scintillators, one with straight modules, two with bended ones. Each layer has a thickness of $5$mm. It is used for a rough determination of the hit position on the higher level and as a starting value for the FPC analysis.
\item \textbf{FRH} The Forward Range Hodoscope\\
The kinetic energy of the particles is measured by the FRH. It consists of $4$ layers of cake-piece shaped plastic scintillators of $11\mathrm{cm}$ thickness. There are $24$ scintillators pro layer. It is also used for particle identification by the dE-E technique.
\item \textbf{FRI} The Forward Range Interleaving Hodoscope\\
Between third and forth layer of FRH two layers of plastic scintillators are installed (FRI). Each layer is made of $32$ strips of $5.2$mm thickness. The FRI is used to determine the scattering angles of neutrons.
\item \textbf{FVH} The Forward Veto Hodoscope\\
The last layer of FD is FVH. It consists of $12$ horizontally oriented scintillator strips with photomultipliers on both sides. The hit position is determined from the time differences of the signals. It is used to identify particles which are not stopped in the FRH. 
\end{itemize}

\subsubsection{The Central Detector}
\label{subsubsec:cd}
The Central Detector (CD) surrounds the interaction point and is constructed to identify energies and angles of the decay products of $\pi^{0}$ and $\eta$ mesons, with close-to $4\pi$ acceptance.  It consists of:
\begin{itemize}
\item \textbf{SCS} The Superconducting Solenoid\\
The SCS produces an axial magnetic field necessary for momentum reconstruction using the inner drift chambers. As superconductor NbTi/Cu is used cooled down by liquid He at $4.5\mathrm{K}$. The maximal central magnetic field is $1.3\mathrm{T}$. The return path for the field is done by a yoke made of $5$ tons of pure iron  with low carbon content.
\item \textbf{MDC} The Mini Drift Chamber\\
The MDC is build around the beam pipe and it is used for momentum and vertex determination \myImgRef{mdc}.  It consists of $17$ layers with in total $1738$ straw tubes detectors. It covers scattering angles from $24^{\circ}$ to $159^{\circ}$ \cite{jacewicz}. For the resolution refer to \myTabRef{tab:mdc}.

\myTable{
\begin{tabular}{|l|c|c|}
\hline
 particle& $p$ $\mathrm{[ MeV/c ]}$ & resolution $\bigtriangleup p/p$ \\ 
\hline
\hline
electrons & $20-600$ & $<1\%$ \\ 
\hline
 pions, muons & $100-600$ & $<4\%$ \\ 
\hline
protons & $200-800$ & $<5\%$\\
\hline
\end{tabular}
}{MDC resolution}{tab:mdc}

\begin{figure}[ht!bp]%
	\begin{center}%
	\frame{
		\includegraphics[height= 0.7\textwidth, angle=270]{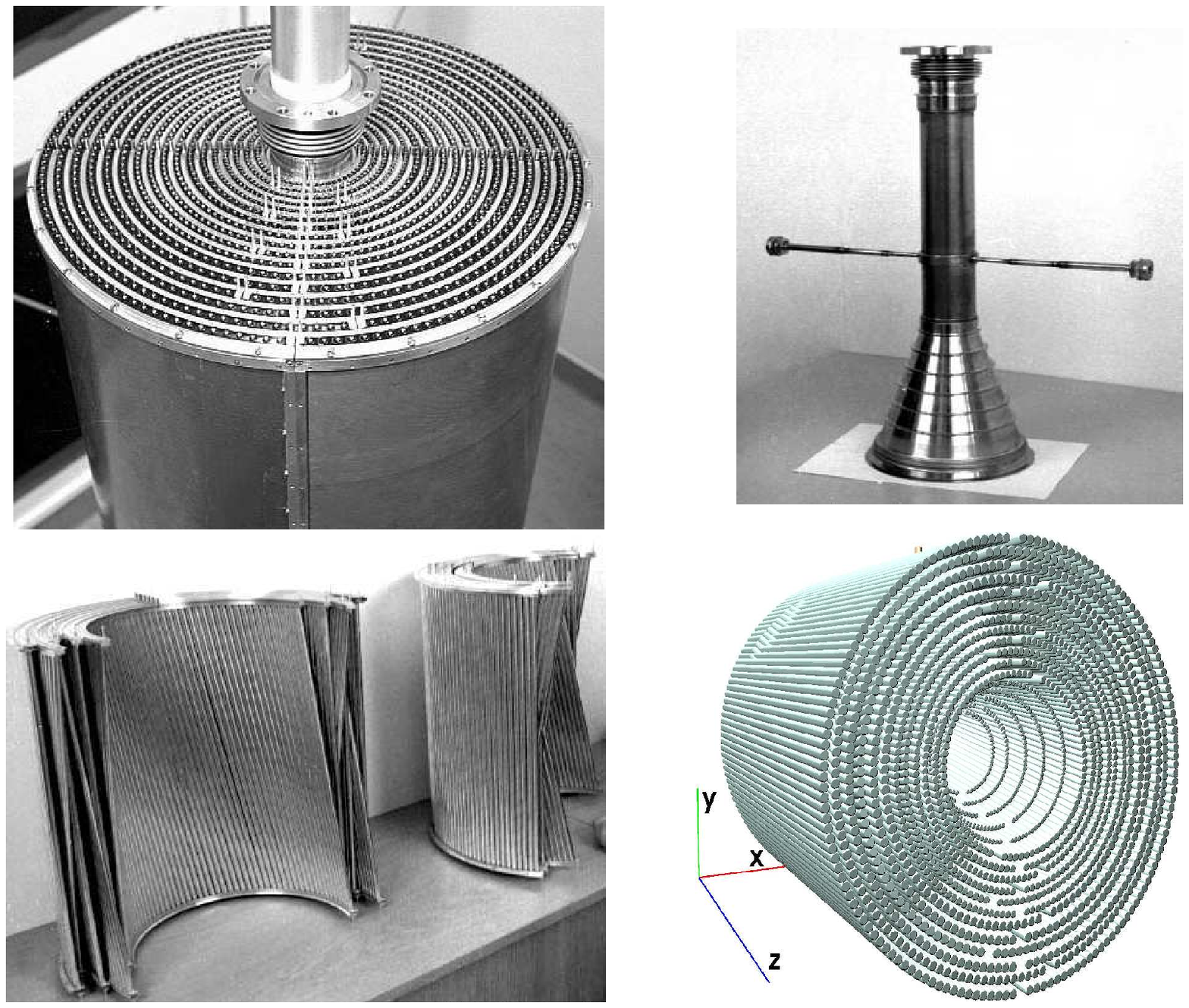}%
		}
		\caption[MDC and Be pipe.]{MDC and Be pipe. The fully assembled MDC inside Al-Be cylinder (upper left)\cite{jacewicz}.}
		\label{image_mdc}%
	\end{center}%
\end{figure}

\item \textbf{PSB} The Plastic Scintillator Barrel\\
The PSB surrounds the MDC inside the SCS. It consists of $146$ $8\mathrm{mm}$ thick strips that form a barrel like shape. It is used together with MDC and SEC, and acts as a dE-E and dE-momentum detector and  as well as a veto for photons.

\item \textbf{SEC} The Scintillator Electromagnetic Calorimeter\\
\label{SECofWASA}
The SEC is the heart of the WASA detector and maybe the most important part.At CELSIUS it has been used to measure electrons and photons up to $800$~MeV. However, using a different setting the energy range can be extended taking into account the higher energy available at COSY. It consists of $1012$ CsI(Na) crystals shaped like a truncated pyramids \myImgRef{crystal}. It covers angles from $20^{\circ}$ to $169^{\circ}$ and the crystals are placed in $24$ layers along the beam \myImgRef{sec2}. The lengths of the crystals vary from $30\mathrm{cm}$ (central part), $25\mathrm{cm}$ (forward part), $20\mathrm{cm}$ (backward part). The forward part consists of $4$ layers each $36$ crystals, covering the range of $20^{\circ}-36^{\circ}$. The central part consists of $17$ layers with $48$ elements each, covering the range between $36^{\circ}-150^{\circ}$, and the backward part with $3$ layers, two with $24$ crystals and one with $12$. The geometrical distribution of the crystals can be seen in \myImgRef{sec1} and \myImgRef{sec3}.

\myFrameSmallFigure{crystal}{CsI(Na) crystal fully equipped with light guide, PM tube and housing \cite{zabierowski}.}{CsI(Na) crystal fully equipped}

\myFrameSmallFigure{sec2}{SEC planar map (arrow indicates beam direction) \cite{proposal}.}{SEC planar map}

\begin{figure}[ht!bp]%
	\begin{center}%
	\frame{
		\includegraphics[height= 0.5\textwidth, angle=270]{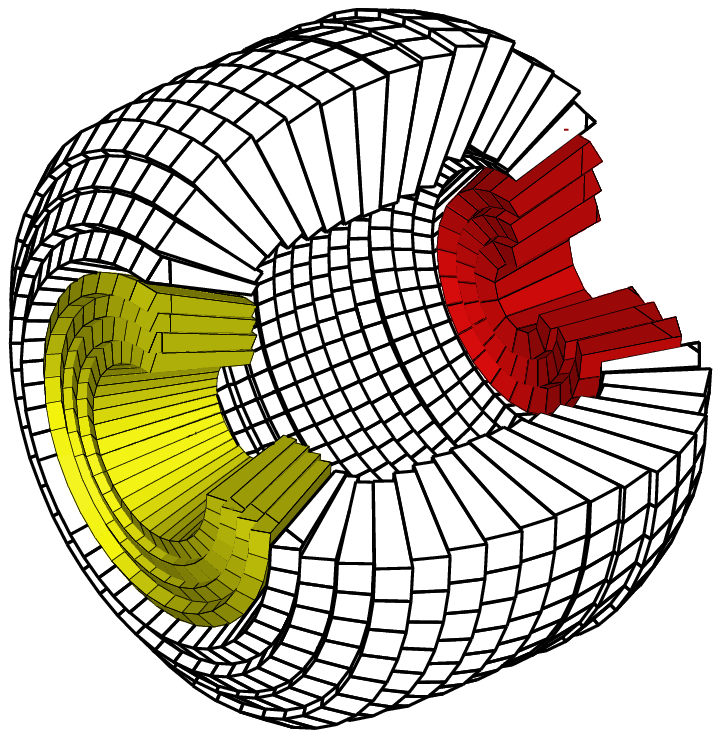}%
		}
		\caption[SEC schematic view.]{Schematic view of the calorimeter layout. It consists of the forward part (yellow, on the left), central part and the backward part (red, on the right) \cite{proposal}.}
		\label{image_sec1}%
	\end{center}%
\end{figure}

\myFrameSmallFigure{sec3}{Photo of the SEC (forward to the left) \cite{wwwWASA}.}{Photo of the SEC}


The calorimeter is composed of sodium doped CsI crystals. They are painted with transparent varnish for moisture protection and wrapped in $150\mathrm{\mu m}$ teflon and $25\mathrm{\mu m}$ aluminized mylar foil \cite{zabierowski}. For more information refer to \myTabRef{tab:sec}.\\
\myTable{
\begin{tabular}{|l|c|}
\hline
 amount of active material& $16$~$X_0$ \\ 
\hline
geometric acceptance:& $96\%$\\
polar angle:&$20^{\circ}-169^{\circ}$\\
azimuthal angle:&$0^{\circ}-180^{\circ}$\\ 
\hline
 relative energy resolution:& $30\%$ (FWHM)\\ 
$Cs(137) 662\mathrm{keV}$ & \\
\hline
maximal kinetic energy for stopping: &\\
pions/protons/deuterons & $190/400/500 \mathrm{MeV}$ \\
\hline
\end{tabular}
}{SEC parameters}{tab:sec}

The advantages of using this type of material instead of CsI(Tl) are the following:
\begin{itemize}
\item emission peak at $420\mathrm{nm}$ (CsI(Tl): $550\mathrm{nm}$) matches very well the spectral efficiency of the most commonly~used~PMT's 
\item the scintillation time is short
\item much less afterglow
\item much more radiation durability
\end{itemize}

\bigskip
Detailed comparison between several types of CsI crystals is presented in \myTabRef{tab:CsIprop}.
\bigskip
\myTable{
\begin{tabular}{|l||c|c|c|}
\hline
Scintillator& CsI & CsI&  CsI \\
(Activator)& (Tl) & (Na) & (undoped)\\
\hline 
\hline
Density [$\textrm{gcm}^{-3}$] & 4.51 & 4.51 & 4.51 \\ 
\hline
Hygroscopic & slightly & yes &  slightly\\ 
\hline
Emission wavelength max [nm] & $550$ & $420$ & $315$ \\ 
\hline
Lower Cut-off [nm]& $320$ &  $300$ &  $260$\\ 
\hline
Refractive index at emission max & $1.79$ &  $1.84$ &  $1.95$\\ 
\hline
Primary decay time [$\mu s$] &  $1.0$ & $0.63$ &  $0.016$\\ 
\hline
Light yield [$10^3$ photons/MeV ] & $52-56$ & $38-44$ & $2$ \\ 
\hline
\end{tabular}
}{Properties of CsI scintillator crystals}{tab:CsIprop}
 
\end{itemize}

\subsection{Physics at WASA}
\label{subsec:wasaPhysics}
The physics program at WASA will investigate symmetries and symmetry breaking as well as hadron structure and interactions. The planned experiments address the following topics: mixing of the scalar mesons $a_{0}/f_{0}(980)$, study of hyperon resonances, $a^{+}_{0}$ production, pentaquarks, isospin violation in $\overrightarrow{d}d\longrightarrow \alpha\pi^{0}$ so it is permeated with newness. 

However the most promising and intriguing problems, that maybe at WASA will be clarified, of the whole physics program are \textbf{rare and very rare decays of $\eta$ and $\eta^{'}$ mesons.} Below I will try to outline the physics behind them.
\bigskip

The $\eta$ and $\eta^{'}$ belong to the SU(3) lightest nonet \myImgRef{nonet} of the pseudoscalar mesons ($0^{-}$) \cite{perkins}. 

\begin{figure}[ht!bp]%
	\begin{center}%
	\frame{
		\includegraphics[height= 0.7\textwidth, angle=270]{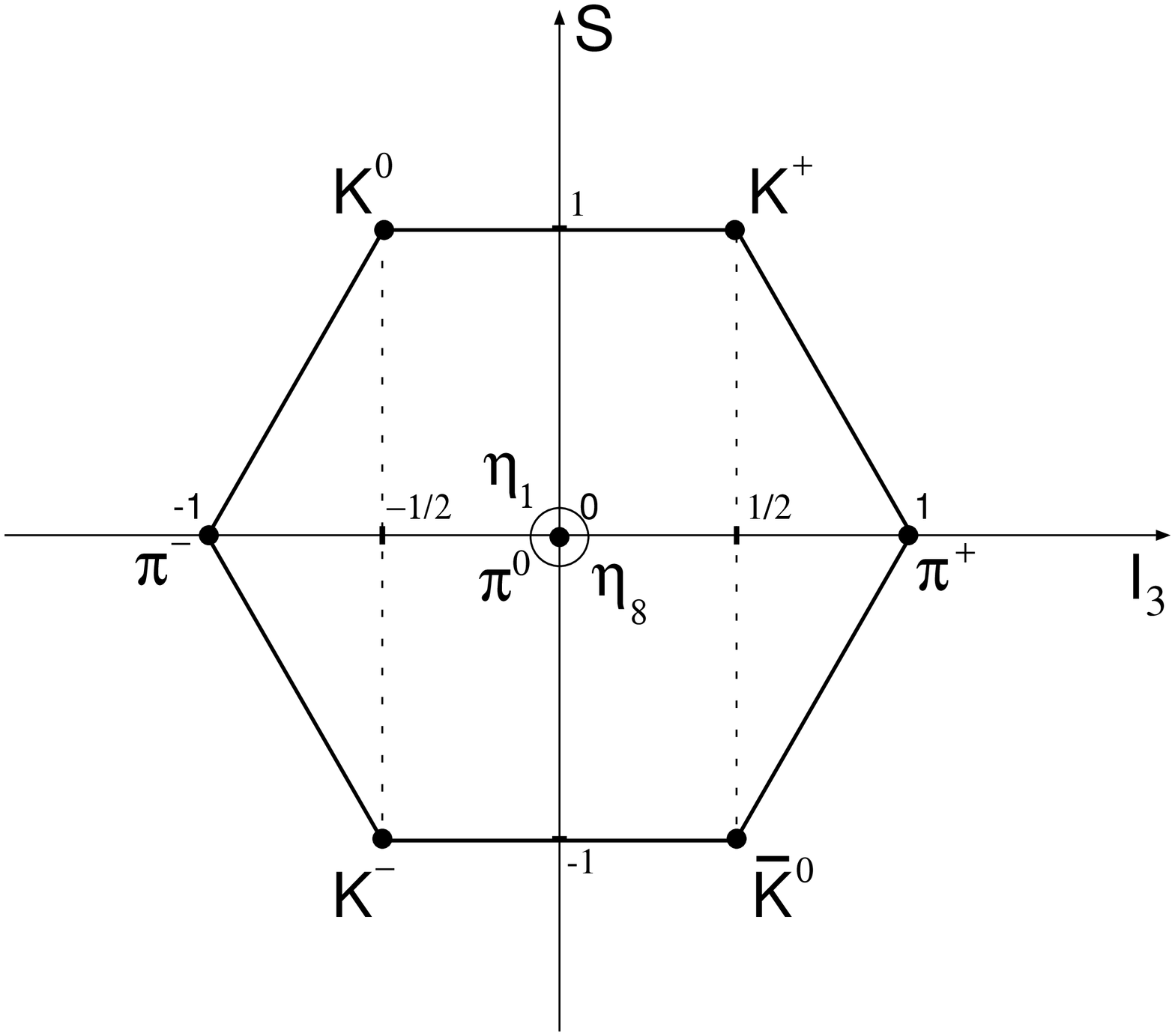}%
		}
		\caption[SU(3) nonet of pseudoscalar mesons]{SU(3) nonet of pseudoscalar mesons.}
		\label{image_nonet}%
	\end{center}%
\end{figure}
We can write the pure SU(3) states as:
\[
\left.\begin{array}{ccc}
      \eta_{8} & = & \frac{1}{\sqrt{3}}(u\bar{u}+d\bar{d}+s\bar{s})\\
			& &\\
     \eta_{1} & = & \frac{1}{\sqrt{6}}(u\bar{u}+d\bar{d}-2s\bar{s})\end{array}\right\} I=0
\]
\[
\left.\begin{array}{ccc}
     \widetilde{ \pi}^{0} & = & \frac{1}{\sqrt{2}}(d\bar{d}-u\bar{u})\end{array}\right\} I=1
\]
States with the same quantum numbers can mix so $\widetilde{\eta}$ and $\widetilde{\eta}^{'}$ are not pure SU(3) states only mixes of the singlet and octet states \cite{moskal}:
\[
\begin{array}{ccc}
      \widetilde{\eta} & = & \cos(\theta_{p})\eta_{8}-\sin(\theta_{p})\eta_{1}\\
			& &\\
      \widetilde{\eta}^{'}& = & \sin(\theta_{p})\eta_{8}+\cos(\theta_{p})\eta_{1}
\end{array}
\]
The mixing angle is rather small $\theta_{p}=-11.5^{\circ}$ \cite{pdg} (different experimental approaches lead to different values between $-10^{\circ}$ and $-20^{\circ}$). Mixing like this is allowed by the charge symmetry. But the situation seems not to be so simple, QCD Hamiltonian
\[
H_{QCD}=H_{Color}+H_{Coulomb}+\underbrace{m_{u}u\bar{u}+m_{d}d\bar{d}+m_{s}s\bar{s}}_{H_{m}}
\]
except the Color part $H_{Color}$ contains also the Coulomb part $H_{Coulomb}$  and the part with quark masses $H_{m}$ which mainly leads to non zero probability of transition between two different isospin states, so the transitions between the $ \widetilde{\eta},  \widetilde{\eta}^{'},  \widetilde{\pi}^{0}$ are possible. The physical mesons $\eta, \eta^{'}, \pi^{0}$ are not pure isospin states but only mixtures. Because of the high mass difference between the $\eta$ and $\eta^{'}$ the strongest mixing effect would be observed for the pair $\pi^{0}-\eta$. Lets now write the real meson states  $\pi^{0}, \eta$:
\[
\begin{array}{ccc}
      \pi^{0} & = & \cos(\theta_{\pi\eta})\widetilde{\pi}^{0}-\sin(\theta_{\pi\eta})\widetilde{\eta}\\
			& &\\
      \eta& = & \sin(\theta_{\pi\eta})\widetilde{\pi}^{0}+\cos(\theta_{\pi\eta})\widetilde{\eta}
\end{array}
\]
where $\theta_{\pi\eta}$ is the mixing angle between $\pi^{0}-\eta$. The easiest way to determine the $\theta_{\pi\eta}$ is to measure the ratio of the widths between two decay modes of the $\eta^{'}$ meson $\Gamma(\eta^{'}\longrightarrow3\pi)$ (not allowed by isospin symmetry) to $\Gamma(\eta^{'}\longrightarrow\eta 2\pi)$ (allowed by isospin symmetry) \myTabRef{tab:decays}:
 \[
\begin{array}{ccc}
       R_{1}& = &\frac{\Gamma(\eta^{'}\longrightarrow\pi^{0}\pi^{0}\pi^{0})}{\Gamma(\eta^{'}\longrightarrow\eta\pi^{0}\pi^{0})} \\
			& &\\
      R_{2}& = & \frac{\Gamma(\eta^{'}\longrightarrow\pi^{0}\pi^{+}\pi^{-})}{\Gamma(\eta^{'}\longrightarrow\eta\pi^{+}\pi^{-})}
\end{array}
\]
and then the $\theta_{\pi\eta}$ can be expressed \cite{etaRef}:
\[
\begin{array}{ccc}
R_{i}&=&P_{i}\sin^{2}(\theta_{\pi\eta})\\
 & &\\
\sin(\theta_{\pi\eta})&=&\frac{\sqrt{3}\bigtriangleup m}{4 (m_{s}-\widehat{m})}
\end{array}
\]
wheres $P_{i}$ is the phase-space factor, $\bigtriangleup m = m_{d}-m_{u}$, $\widehat{m}=(m_{d}+m_{u})/2$.
\bigskip

Not only these four decay modes will be investigated, thanks to the high production rate \myTabRef{tab:prod}, WASA opens possibility to study really \textbf{very rare} processes \myTabRef{tab:decays}. Lots of these reaction will be measured first time in the world. For more details look at \cite{proposal}. 
\bigskip
\myTable{
\includegraphics[]{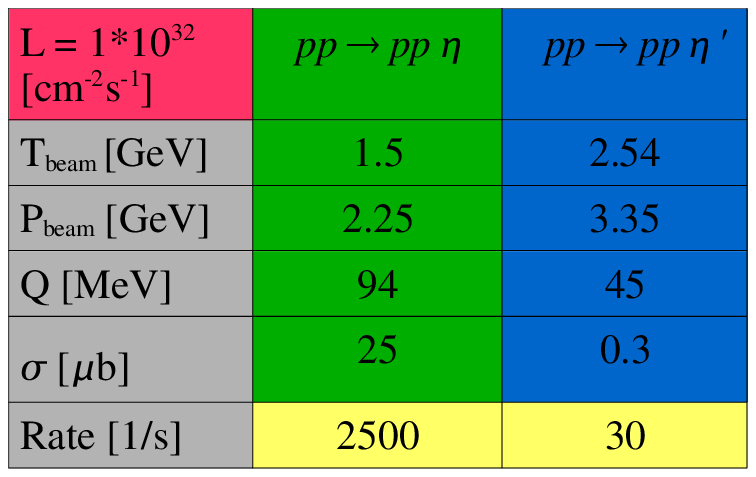}
}{Production rates at WASA}{tab:prod}
\myTable{
\includegraphics[width=0.7\textwidth]{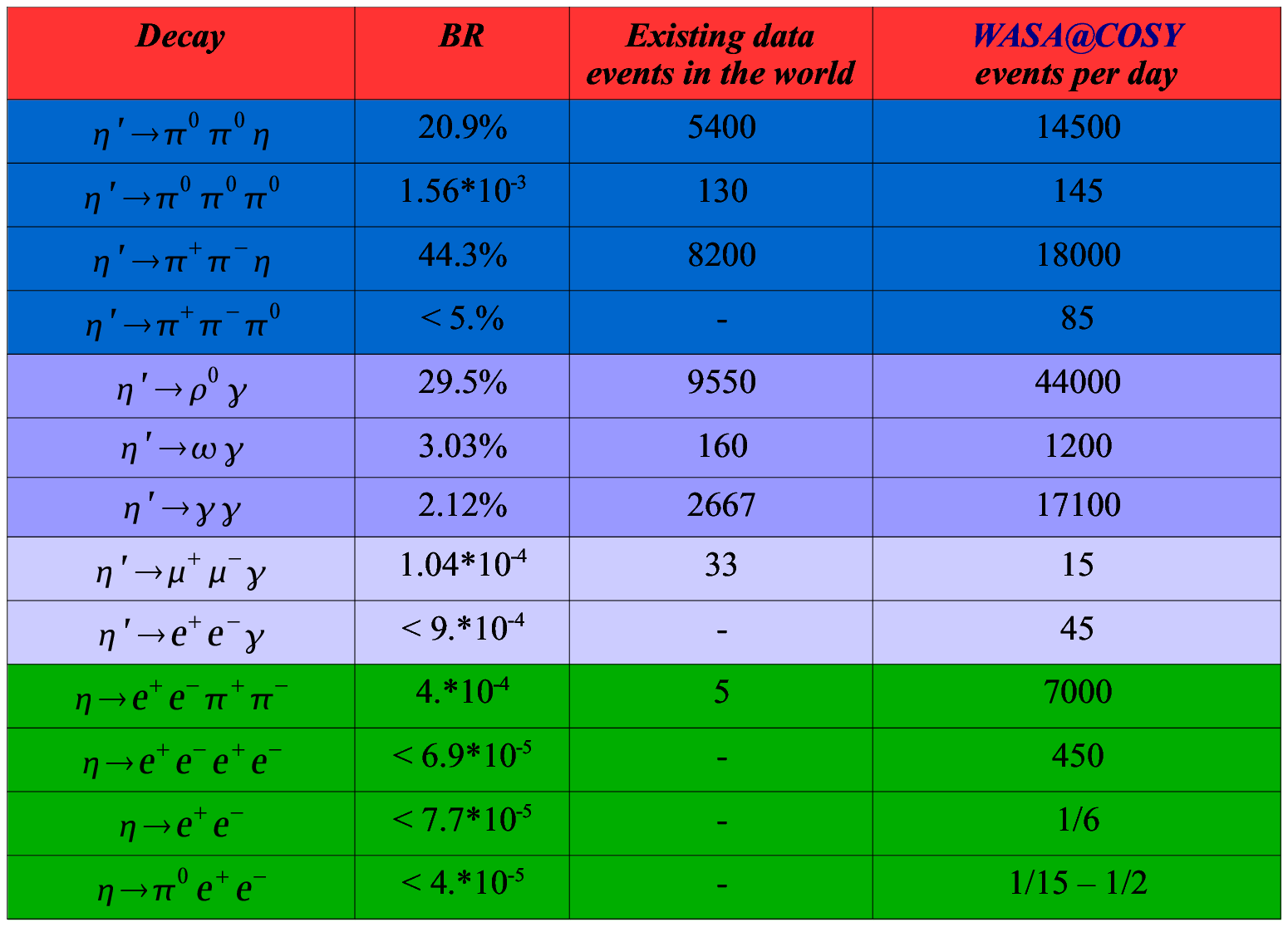}
}{Decays of $\eta$ and $\eta^{'}$ at WASA}{tab:decays}
\emptydoublepage
\section{Physics of Calorimetry}
\label{sec:calorimetry}

Calorimeters are devices used to measure energy of electrons, photons and hadrons. In order to do this they are usually build as a block of instrumented material in which the particles are fully absorbed and their energy is transformed to the measureable quantity. 
The calorimeters could be divided in general into hadronic and electromagnetic calorimeters. Hadronic calorimeters are used to measure hadrons through their strong and electromagnetic interactions. Due to involvement of nucleon (hadronic) interactions the energy deposit is widely spread and the resolution is quite bad (typically $30-50\%$). 
Electromagnetic calorimeters are used to measure energy and position of electrons(positrons) and photons by means of their electromagnetic interactions (bremsstrahlung, pair production) with matter.
I will concentrate on the latter ones i.e. \textit{the~electromagnetic~calorimeters}. 
To understand how they work first the electromagnetic interactions will be investigate.
\subsection{Some definitions}

\myFrameSmallFigure{enloss1}{Energy loss of electrons per radiation length in Pb as a function of energy \cite{pdg}.}{Energy loss of electrons per radiation length in Pb as a function of energy}

\myFrameSmallFigure{phcr1}{Photon interaction cross section in Pb as a function of energy \cite{cal}.}{Photon interaction cross section in Pb as a function of energy}

Electrons and photons interact with matter over today good understood QED processes, so the description of them is possible using simple empirical approach. There are two main regimes that govern their interactions  \myImgRef{enloss1} shows the fractional energy loss of electrons in lead and \myImgRef{phcr1} shows photon interaction cross section.  As can be seen for energies larger than $\sim10\textrm{MeV}$ photons mainly interact by pair production, while electrons lose their energy mostly by bremsstrahlung. For energies $>1\textrm{GeV}$ these processes become energy independent. In the lower energy regime, for photons compton and photo-electric effect dominate, for electrons however ionization rises, traversing electron or positron in matter undergoes collisions with molecules and atoms. It is considered as a ionization when energy loss per collision is below $0.255$MeV. Although other processes also contribute: Bhabha scattering and annihilation of positrons and M\"oller scattering for electrons. The consequence of this behaviour is that the sufficiently high energetic photons or electrons interact with material producing secondary electrons by pair production and photons by bremsstrahlung. These secondary particles produce again further particles giving rise for an\textit{ electromagnetic shower} \myImgRef{cascade1}.
This process continues as long as the energy is high enough to produce other particles. The limit is indicated by so called \textit{critical energy} $E_c$ (to be defined later). 
\myFrameSmallFigure{cascade1}{Possible shower development induced by few MeV gamma, the energies of $e^{\pm}$, $\gamma$ are given in keV. Following processes are indicated: A~-~conversion to $e^{+}e^{-}$ pair, B~-~annihilation of $e^{+}$ into two gammas, C,~D,~F~-~Compton scattering, E,~G~-~photoelectric effect \cite{cal2}.}{Possible shower development induced by few MeV gamma}

The propper scale for the shower process (expansion description) is the \textit{radiation length} $X_0$, usually given in gcm$^{-2}$\cite{pdg}. That is defined as:
\begin{itemize}
\item mean distance over which the electron will lose $1/e$ of its initial energy by bremsstrahlung
\begin{equation}
\langle E(x) \rangle=E_0\exp\left[-x/X_0\right],
\label{radlE}
\end{equation} 
\item 7/9 of the mean distance over which the photon beam will lose $1/e$ of its intensity 
\begin{equation}
\langle I(x) \rangle=I_0\exp\left[-x7/(9X_0)\right].
\label{radlPh}
\end{equation} 
\end{itemize}

The \textit{radiation length} depends on the type of material, and can be expressed as \cite{pdg}:
\begin{equation}
X_0=\frac{716.4\, \textrm{gcm}^{-2}A}{Z(Z+1)\ln(287/\sqrt{Z})},
\label{radlMaterial}
\end{equation}
where $Z$~--~atomic number of the material, $A$~--~weight number of the material.

In case of compound material it can be approximated by:
\begin{equation}
1/X_0=\sum w_j/X_j,
\label{radlApprox}
\end{equation}
where $w_j$~--~fraction by weight of the j-th element, $X_j$~--~radiation length of the j-th element.
\bigskip

\myFrameSmallFigure{ec2}{Two definitions of the critical energy $E_c$ \cite{pdg}.}{Two definitions of the critical energy}

The \textit{critical energy} $E_c$ mentioned above is not so clear-cut defined. There are two definitions used \myImgRef{ec2}:
\begin{itemize}
\item \textit{critical energy} is the energy at which ionization energy loss of electron is equal energy loss by bremsstrahlung
\begin{equation}
\left.\frac{dE}{dx}\right\vert_{brems}=\left.\frac{dE}{dx}\right\vert_{ioniz}
\label{Ecrit1}
\end{equation}
\item \textit{critical energy} is the energy at which ionization energy loss equals electron energy $E$ divided by radiation length $X_{0}$ 
\begin{equation}
\left.\frac{dE}{dx}\right\vert_{ioniz}=\frac{E}{X_0},
\label{Ecrit2}
\end{equation}
this is the same as the first definition \myEqRef{Ecrit1} with approximation
\begin{equation}
\left.\frac{dE}{dx}\right\vert_{brems}\thickapprox\frac{E}{X_0}.
\label{Ecriapprox}
\end{equation}
\end{itemize}
\bigskip

As shown in \myEqRef{radlE} and \myEqRef{radlPh} the scale for electromagnetic cascade developed by incident photon or electron is the same and is expressed by radiation length $X_0$. So now we can describe in proper way the dimensions of the cascade.
\myFrameFigure{profile1}{Longitudinal shower profiles simulated in $\mathrm{PbWO_{3}}$ in dependence of the material thickness in radiation length, for $e^{-}$ energy $1\mathrm{GeV}$, $10\mathrm{GeV}$, $100\mathrm{GeV}$, $1\mathrm{TeV}$ from left to right \cite{cal}.}{Longitudinal shower profiles simulated in $\mathrm{PbWO_{3}}$}

The \textit{mean longitudinal shower profiles} of energy deposits \myImgRef{profile1} are well described by a gamma distribution:
\begin{equation}
\frac{dE}{dt}=E_0 \frac{b^{a}{t}^{a-1}e^{-bt}}{\Gamma(a)},
\label{gammaProfile}
\end{equation}
where $E_0$ is the incident particle energy, $a, b$ are parameters depending on the type of incident particle ($e^{\pm}$, $\gamma$) and $t$ is depth in the material in terms of radiation length. The maximum of this distribution is located at $t_{max}=(a-1)/b$ which is usually estimated as:
\begin{equation}
t_{max}=\ln\left(\frac{E_0}{E_c}\right)+t_{0},
\label{tMAXestim}
\end{equation}
where $t_{0}$ depends on type of the particle and is $-0.5$($0.5$) for an electron(photon) induced cascade.($E_{c}$ is critical energy defined earlier.)\\
It can be seen in \myEqRef{tMAXestim} that the expansion of the electromagnetic cascades is logarithmic and in consequence the detector thickness needed to absorb the energy scales the same way. The thickness that contains the $95\%$ of the shower energy , for material with atomic number $Z$, can be derived as:
\begin{equation}
t_{95\%}=t_{max}+0.08Z+9.6.
\label{t95}
\end{equation}
The transverse development of the shower is mainly caused by multiple scattering of $e^{\pm}$ away from the shower axis. Bremsstrahlung photons of these $e^{\pm}$ also contribute to the cascade spread. The \textit{transverse profile of the shower} is usually characterised by \textit{the Moli\'ere radius} $R_M$. Transverse size integrated over full shower depth, representing the mean lateral deflection of electrons after traversing one radiation length and it is expressed as:
\begin{equation}
R_M=E_s\frac{X_0}{E_c},
\label{Rmoliere}
\end{equation}
with $E_s$ scale energy $\sqrt{4\pi/\alpha}m_{e}c^2\thickapprox21\textrm{MeV}$ (here $\alpha$ Fine structure constant) and $E_c$ critical energy from \myEqRef{Ecrit2}. On the average in a cylinder of $\thicksim1R_M$ radius $90\%$ of shower energy is contained. The transverse size of the electromagnetic shower is energy independent as can be seen on the \myImgRef{radial2}.

\myFrameFigure{radial2}{Transverse shower profiles simulated in $\mathrm{PbWO_{3}}$ in dependence of the transverse distance from the shower axis in radiation length's for $e^{-}$ energy $1\, \mathrm{GeV}$(closed circles) and $1\, \mathrm{TeV}$(open circles) \cite{cal3}.}{Transverse shower profiles simulated in $\mathrm{PbWO_{3}}$}

\subsection{The electromagnetic calorimeter itself}
The principle of energy measurement with an electromagnetic calorimeter is that the energy deposited in calorimeter medium by charged particle of the cascade is proportional to the energy of the incident particle. Therefore the total track length of the cascade $T_0$(sum of all ionization tracks in cascade) is proportional to
\begin{equation}
T_{0} \thicksim X_0 \frac{E_0}{E_c},
\label{trackCascade}
\end{equation}
where $E_0/E_c$~--~number of particles in the cascade. So the measurement of the signal produced by particles in the shower gives estimation of the incident particle energy.

As an example of the calorimeter let us consider the electromagnetic calorimeter of the WASA detector setup(~section~\ref{subsubsec:cd}~on~page~\pageref{SECofWASA}~). It is a scintillation calorimeter with inorganic CsI(Na) crystals. The mechanism of scintillation is related to the crystalic structure of the material. Charged particles are producing electron-hole pairs in the conduction and valence bands of the medium. When electrons return to the valence band the photons are emitted. The responce time and the wave length of the photons depends on the gap between the valence and conduction band and on the electron transport in the lattice structure. To increase the light yield and the response time crystals are usually doped with some amounts of inpurities, in our example Na, that create additional activation sites in gap between the bands thus increasing the probability of light emission.\\
So in this case the amount of light, emitted by scintillation process, induced by electromagnetic shower and collected by photomultiplier is proportional to the energy of incident particle.

The \textit{energy resolution} of ideal calorimeter (infinite size, homogeneous response) is govern mainly by track length fluctuations. The shower development, as known, is a stochastic process, thus the energy resolution:
\begin{equation}
\sigma(E) \thicksim \sqrt{T_0},
\label{resolMain}
\end{equation}
using \myEqRef{trackCascade} we can derive:
\begin{equation}
\frac{\sigma(E)}{E} \thicksim  \frac{1}{\sqrt{T_0}} \thicksim \frac{1}{\sqrt{E_0}}.
\label{resolutionMain2}
\end{equation}

The energy resolution of a realistic calorimeter is not so simple defined. Usually it is written the following way:
\begin{equation}
\frac{\sigma(E)}{E} =  \frac{a}{\sqrt{E}} \oplus \frac{b}{E} \oplus c,
\label{resolutionReal}
\end{equation}
where $\oplus$~--~addition in quadrature.
\bigskip

The three terms in \myEqRef{resolutionReal} contribute into resolution:

\begin{itemize}
\item \textit{the stochastic term} $a/\sqrt{E}$\\
This term comes from the internal fluctuations of shower development, as described above. Typical values for this contribution are on the level of few percents.

\item \textit{the noise term}  $b/E$\\
This term comes from electronic noise of readout chain and depends on features of the circuit. This contribution to energy resolution increase with decreasing energy and may be dominant for energies below few GeV.

\item \textit{the constant term }$c$\\
This term includes effects that do not depend on the incident energy of the particle, such as instrumental effects like imperfection of mechanical structure, temperature gradients, detector aging, radiation damage etc. All these effects cause nonconformities of the response signal, thus smearing measured energy. Usually this term should be kept in orders of $1\%$.

\end{itemize}

Sample energy resolution curves fitted by \myEqRef{resolutionReal} are presented in \myImgRef{tapsres1} for three different scintillator crystals. The parameters of fit are for $\mathrm{CeF_{3}}$ $a=2.17\%$, $c=2.7\%$; for $\mathrm{PbWO_{4}}$ $a=1.41\%$, $c=0.9\%$ and for $\mathrm{BaF_{2}}$ TAPS spectrometer $a=0.59\%$, $c=1.9\%$.

\myFrameFigure{tapsres1}{Energy resolution of $\mathrm{CeF_{3}}$ and $\mathrm{PbWO_{4}}$ crystals in comparison with $\mathrm{BaF_{2}}$ scintillators of the TAPS spectrometer \cite{taps}.}{Energy resolution curve for scintillator crystals}

\emptydoublepage
\section{Measurements of the WASA calorimeter \label{sec:data}}
The aim of these measurements was to check the Electromagnetic Calorimeter of WASA after the transport to IKP and to find broken elements as well as those with bad performance. In addition the test results will be used for the relative energy calibration of the CsI(Na) crystals. All channels were checked using source providing $4.4\textrm{MeV}$ photons (~section~\ref{subsec:sourceMeas}~). The results were then compared to cosmics measurements which have been performed for few channels (~section~\ref{subsec:cosmics}~).

\subsection{Electronics setup}
\label{subsec:electronics}
The electronics and readout system used for the test measurements is shown in \myImgRef{circuitE}. In all measurements only one crystal was readout at once. The signals from the crystal \myImgRef{signal} were then amplified by a factor of $100$ [ps~$776$] and split into two branches. One branch was used to construct the trigger: the signal was connected to the discriminator [LeCroy~$821$] in order to adjust the minimum signal level at which the signal should be sampled. The (logical) output of the discriminator was used to create a $12.2\mathrm{\mu s}$ long gate [dual~timer~C.A.E.N.~N$93$B], which defines the time window, in which FlashADC samples the input signal.
The second line from the amplifier was delayed by $600\mathrm{ns}$ (in order to compensate the time for trigger generation and to start sampling well ahead of the signal).
The used Flash~ADC [SIS~3300] had a frequency of $100$~MHz. After digitization the data were sent via crate controller to the computer by a special PCI card and read using data aquisition delivered by ZEL\footnote{\url{http://www.fz-juelich.de/zel}}(EMS system). The data were written to disk. During the measurements the spectra were monitored on-line using the RootSorter\cite{RootSorter} analysis software. 

\myFrameHugeFigure{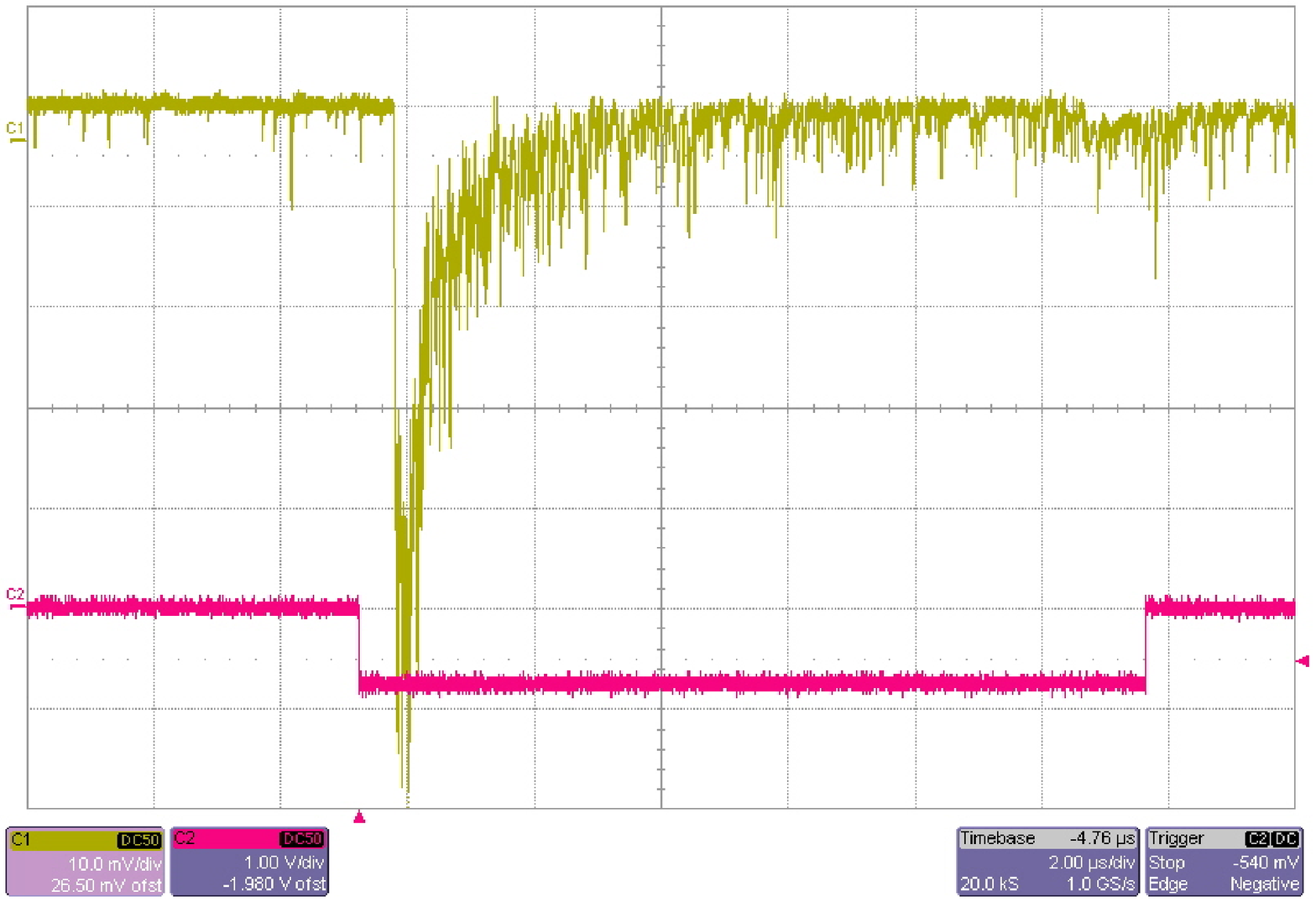}{Raw FlashADC signal from CsI crystal as seen on the oscilloscope. The time gate used is also presented (lower line).}{Raw signal from CsI crystal}
\myFrameHugeFigure{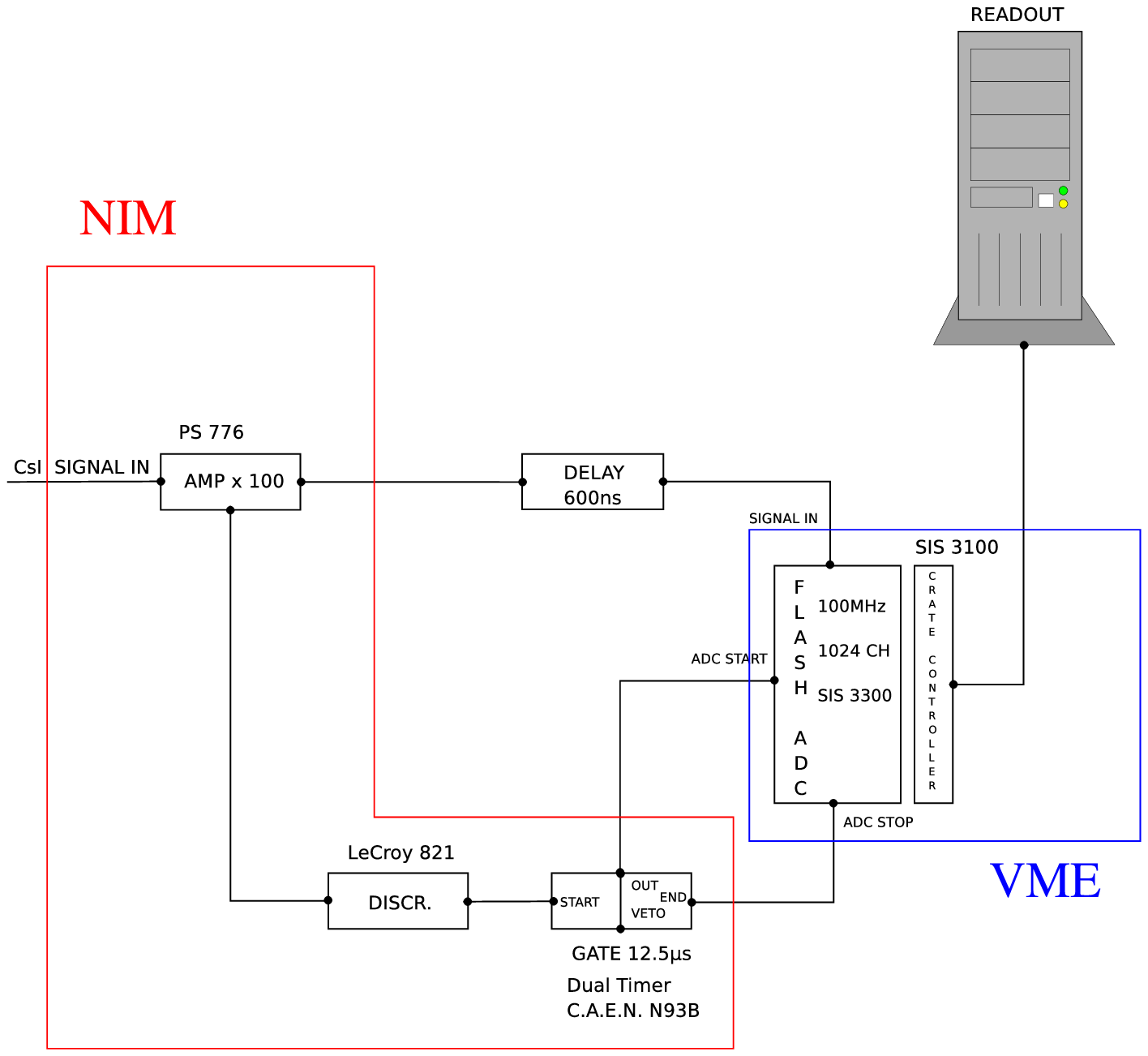}{Electronics setup for measurements}{Electronics setup for measurements}
\subsection{General Analysis}
\label{subsec:generalAnl}
The data were analysed event-by-event. For each sampled pulse \myImgRef{signal} the zero level was computed as a mean value of first $30$ channels and subtracted from the original signal. The integration gate was selected by comparing different spectra using various integration gates \myImgRef{int2}. The signal to background ratio was taken into account as well as the  resolution and the optimal gate was selected. The further analysis was then based on the calculation of the normalized integral within this gate (channel $30$ to $200$). 


\myFrameFigure{signal}{Sample CsI signal with subtracted ground level. The finally selected integration gate marked channel ($30-200$).}{Sample CsI signal with subtracted ground level}
\myFrameHugeFigure{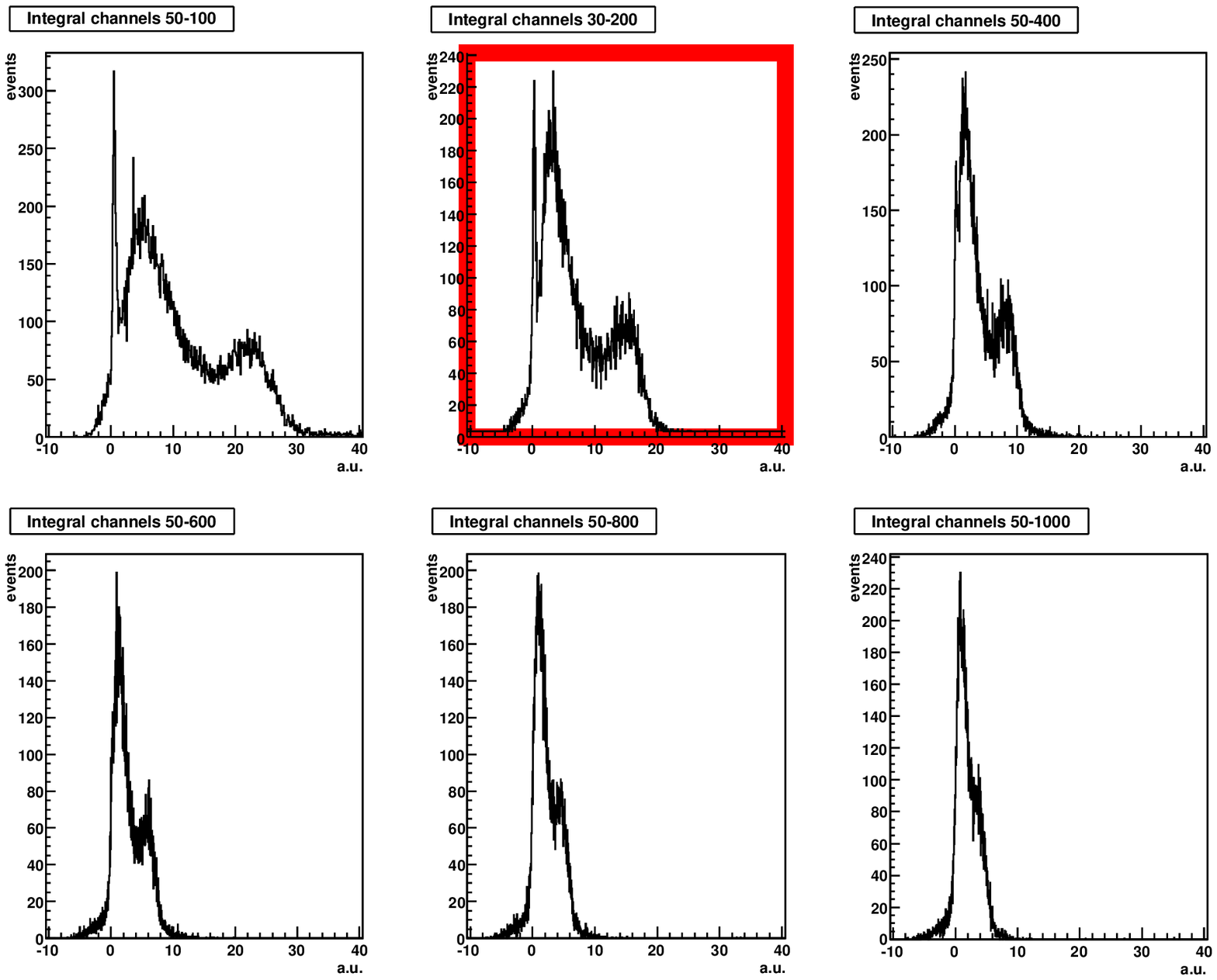}{Integrated pulses for different integration gates. Selected gate is marked.}{Integrated pulses for diffrent integration gates}
\subsection{Source measurements}
\label{subsec:sourceMeas}
\myFrameFigure{ambes1}{Gamma ray spectra of $^{239}\textrm{PuBe}$ source (upper spectrum) and  $^{241}\mathrm{AmBe}$ source (lower spectrum) measured using Na(Tl) scintillator \cite{source}.}{Gamma ray spectra of $^{239}\textrm{PuBe}$ source and  $^{241}\mathrm{AmBe}$ source}

The measurements of individual crystals were done with $^{241}\mathrm{AmBe}$ radioactive source, see \myImgRef{ambes1}, which emits $4.4\textrm{MeV}$ photons. The experimental setup is shown in \myImgRef{schema1} and \myImgRef{stest}. The source was pushed as close as possible to the surface of the CsI crystals. The electronics setup was as described in section~\ref{subsec:electronics}. The data first were treated as explained in section~\ref{subsec:generalAnl}, the spectrum \myImgRef{adcSpectra} is a result of it. The analysis using ROOT\cite{root} was performed, by fitting two functions to the spectra, one describing the background(as exponential) and one describing the signal(as a Gauss function). The exponential background is probably a consequence of electronic noise. Using this technique the peak position and the width of the signal was extracted from the data for each tested module. A typical spectrum with fitted background and signal function is shown in \myImgRef{fitspectra2}.

The peak position of the signal was plotted versus photomultiplier high voltage \myImgRef{hvf3} to check working conditions. Corrections were applied to high voltage setting of the modules to adjust the gain to the same working level. In total around $100$ channels were corrected, the resulting distribution is shown in \myImgRef{pphva}. The full results from performed tests, for each channel, can be find in the Appendix~\hyperlink{appendix}{A} on page~\pageref{appendix}.
\medskip
\myFrameFigure{schema1}{Setup for source measurements.}{Setup for source measurements}
\myFrameFigure{stest}{Photo of setup for source measurements.}{Photo of setup for source measurements}

\myFrameSmallFigure{adcSpectra}{A sample ADC spectrum. The photon peak is indicated.}{A sample ADC spectrum.}
\myFrameSmallFigure{fitspectra2}{One fitted ADC spectrum.}{One fitted ADC spectrum.}
\myFrameSmallFigure{hvf3}{Measured peak position of the crystals versus high voltage.}{Peak position versus high voltage}
\myFrameSmallFigure{pphva}{Measured peak position of the crystals versus high voltage after high voltage adjustment.}{Peak position versus high voltage after adjustment}

\newpage
The peak width $\frac{\sigma}{mean}*100\mathrm{\%}$ for measured crystals was also calculated and is presented in \myImgRef{wpeak} as a histogram. The histogram is centered around the value of $15\%$. The peak width is a convolution of the experimental resolution as well the gamma spectrum distribution from used source \myImgRef{ambes1}. The extraction of real crystal resolution is hard to estimate, but the calculated value is quite satisfactory.
\medskip
\myFrameFigure{wpeak}{Histogram of measured width (in terms of $\sigma$) for different crystals.}{Measured SEC peak width}

\subsection{Cosmics measurements}
\label{subsec:cosmics}
The source tests were supplemented by measuring the response for cosmic muons for a few modules. The electronics setup was the same as described in section~\ref{subsec:electronics}.
Five modules at diffrent azimuthal angles ($0^{\circ}, 45^{\circ}, 90^{\circ}, 135^{\circ}, 180^{\circ}$) have been selected \myImgRef{cosm}. The data were analysed as described in section~\ref{subsec:generalAnl}.

The GEANT3 based Wasa Monte-Carlo program was used to simulate the shape and the peak position of the energy deposits.
Consequently, different geometries and sizes of the crystals were taken into account \cite{inken}. The muons were generated accordingly with a zenith angular distribution of $\cos^{2}(\phi)$ for $\phi=0^{\circ}-80^{\circ}$.
 
\myFrameSmallFigure{cosm}{The positions of the tested crystals in the calorimeter.}{The positions of the tested crystals in the calorimeter}
\myFrameSmallFigure{cosMC}{Simulated Monte-Carlo spectra for cosmic muons in five different crystals.}{Cosmics Monte-Carlo}
\myFrameSmallFigure{mc2}{Simulated Monte-Carlo spectra: upper picture shows crystals at angles of $0$ and $180$, the lower picture at angles of $45$ and $135$.}{Cosmics Monte-Carlo comparison}

The simulated spectra for the five different crystals are presented in \myImgRef{cosMC}. The change of the slopes of tails of distributions is expected and can be reproduced by the simulation. This effect is due to the angular orientation of the measured crystal: most of the cosmics are coming from $\phi=0^{\circ}$, so for different orientations they penetrate different amount of material. The symmetry of the distributions should be visible between the crystals oriented opposite the $90^{\circ}$ direction. The \myImgRef{mc2} shows simulated spectra for crystals oriented opposite to the $90^{\circ}$ axis. 
\myFrameHugeFigure{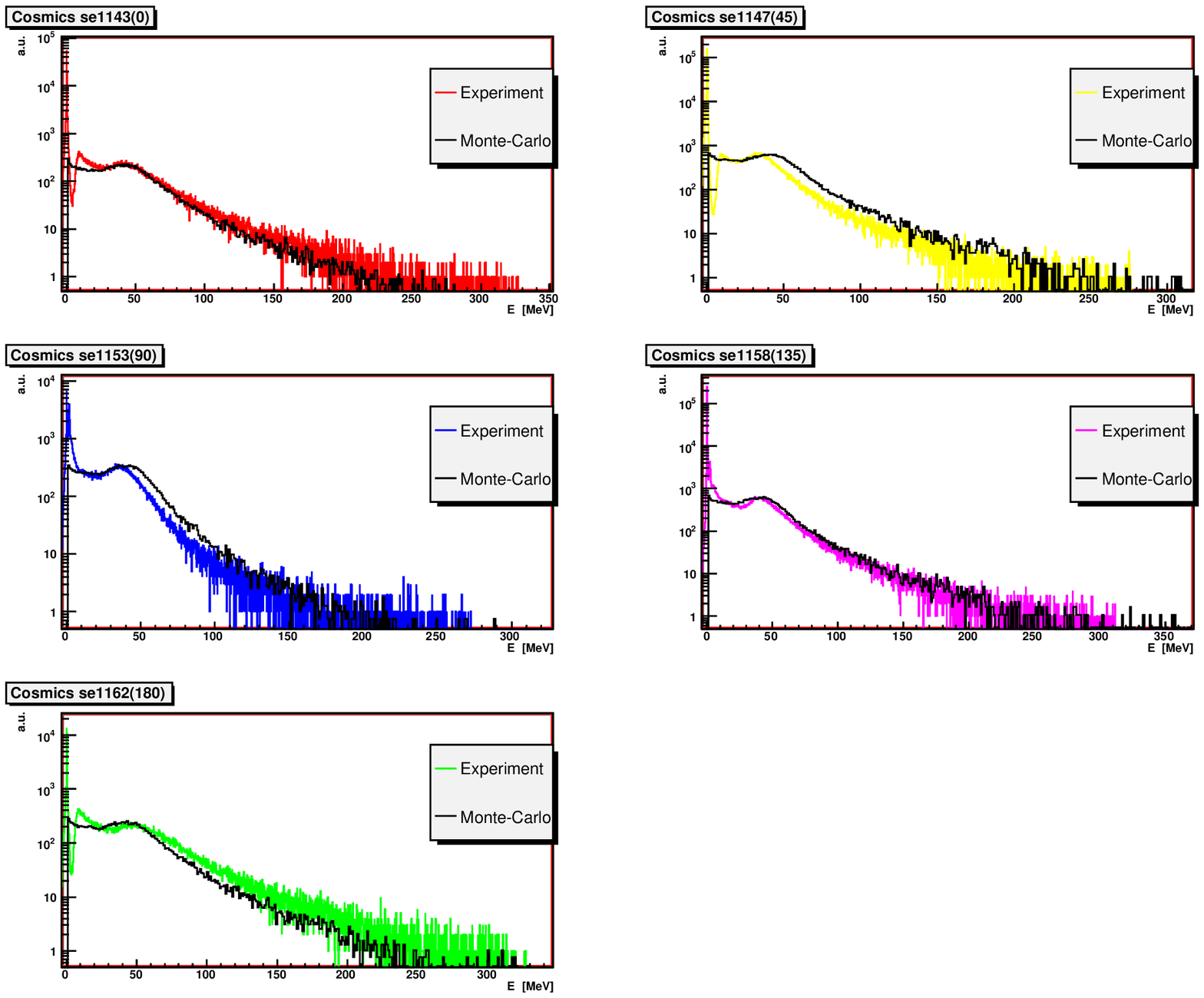}{Measured cosmic muons spectra (calibrated by the $4.4$MeV gammas) compared to the Monte-Carlo spectra.}{Cosmics Measured versus Monte-Carlo -- source calibration}

The comparison was done to check to what extent the source measurements can be used for calibrating the calorimeter. The measured spectra were calibrated using the peak position from source measurements (~section~\ref{subsec:sourceMeas}~) and compared to the simulated ones \myImgRef{MCvsExpCalS}. As one can see, the peak position of the distributions differs: the calibration performed in such a way has a precision of about $\sim10\%$. The explanation of this effect is as follows: The shower induced by $4.4\textrm{MeV}$ gammas has its maximum in the crystal at depth of $\sim3\textrm{cm}$. The crystals are $\sim30\textrm{cm}$ long, so the scintillation light has to pass most of the crystal lenght, reflecting from the aluminized mylar foil wrapping and atenuating. In case of cosmic muons, the crystal is iluminated from all directions with different intensities, the cosmics are minimum ionizing particles. They penetrate the crystal loosing small amount of their energy. This is the reason of diffrences in peak positions. The precision of arrangement of the source in front of the crystal is also important.

\myFrameSmallFigure{cosExpCalC}{Measured cosmic muons spectra calibrated using the simulated peak position.}{Cosmics Measured -- Monte-Carlo calibration}
\myFrameSmallFigure{exp2}{Measured cosmic muons spectra calibrated with simulation: upper picture shows crystals at angles of $0$ and $180$, the lower picture at angles of $45$ and $135$.}{Cosmics measurement comparison}

The next step was to check the consistency of the data, to see the expected changes of the slope of the tails. The measured spectra were thus re-calibrated using the peak position from the Monte-Carlo simulation and compared accordingly \myImgRef{cosExpCalC}. In order to see also the symmetry effect between the spectra for crystals oriented opposite the $90^{\circ}$ direction, \myImgRef{exp2} was plotted. The distributions agree quite well.

To validate also the accuracy of the Monte-Carlo simulation predictions with the performed experiment, the data calibrated using simulation were plotted together with calculated Monte-Carlo spectra for five different crystals respectively \myImgRef{MCvsExpCalC}, the normalization for peak position was applied for each. One can see the spectra are matching very good each other, the shape is described quite satisfactory, from the peak position up to the tails. The conclusion is that the performed Monte-Carlo simulation for the cosmic muons reproduces well the energy deposits in the CsI(Na) crystals.

\myFrameHugeFigure{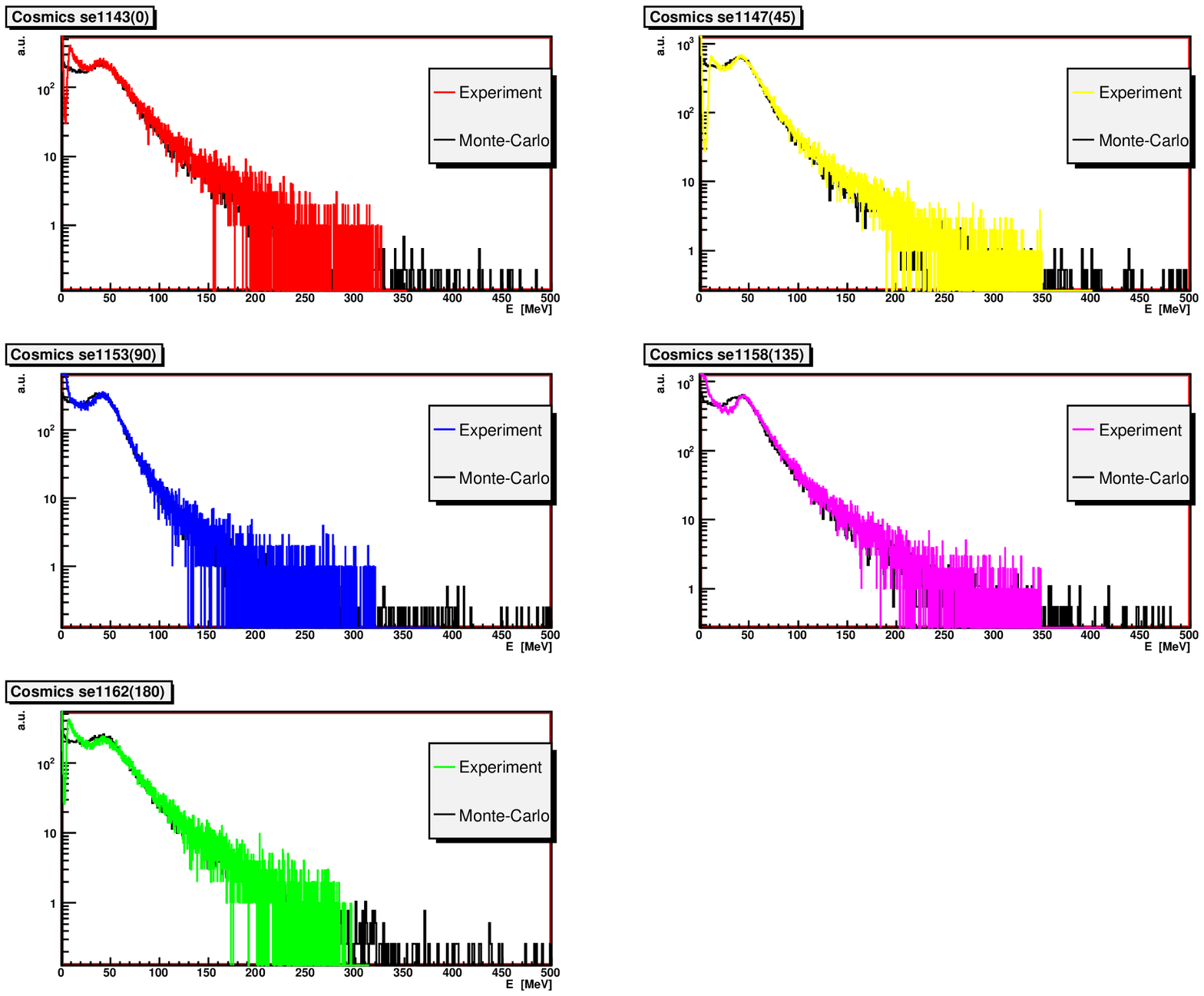}{Measured cosmic muons spectra (calibrated by the peak position from simulation) compared to the Monte-Carlo spectra.}{Cosmics Measured versus Monte-Carlo -- Monte-Carlo calibration}
\emptydoublepage
\section{Summary and conclusions}
\label{sec:summary}
The tests of the electromagnetic calorimeter components (CsI(Na) crystals) for the WASA~at~COSY setup were performed. All available modules were tested to check their properties after transport. Scanning of the modules was performed with a radioactive source providing $4.4\, \mathrm{MeV}$ gammas. Some broken channels were found, fixed or replaced. The energy calibration constant was calculated for each crystal and the peak width for this energy was extracted (in average $15\%$). This width is a consequence of the experimental resolution as well the used gamma source spectrum.
\medskip

\myFrameFigure{fzupp3}{Comparison between peak the positions measured in Uppsala and in Fz-J\"ulich.}{Peak position Uppsala versus J\"ulich}

Similar measurements with a source of the same type were done earlier at TSL in Uppsala before moving WASA to IKP at Fz-J\"ulich \cite{david}. The comparison between the measured peak positions is presented in \myImgRef{fzupp3}. In general, strong correlation between these two measurements is observed, but however, in few cases a discrepancy can be seen. The latter may be the result of changed optical contacts between the light guide and the photomultiplier as an effect of the transport. The data are consistent among the measurements, which proofs the same unchanged good quality of the components after the transfer from the CELSIUS ring to the new site. 

In addition the task of investigating the energy calibration was done using cosmic muons (see~section~\ref{subsec:cosmics}). The measured cosmic spectra were compared with the Monte-Carlo simulation. The experimental bias for calibration was determined from this comparison and estimated on the level of $\thicksim10\%$. 
The WASA calorimeter is in satisfactory condition, as it was in the TLS, and it is ready for mounting at the COSY ring and for commissioning in August. \myImgRef{secresult} shows the invariant mass of two gammas reconstructed in the electromagnetic calorimeter versus the missing mass of two protons extracted from the forward detector. As seen on the right, peaks of $\pi$ and $\eta$ mesons are identified with good energy resolution, which confirms the detection capabilities of WASA. The performed tests of calorimeter prove the feasibility of experiments planned for WASA~at~COSY starting at the beginning of 2007.
\bigskip
\myFrameFigure{secresult}{Left picture: Invariant mass of two gammas reconstructed in the calorimeter versus the missing mass of two protons extracted from the forward detector. Right picture: Invariant mass of two gammas with a cut on missing mass; pion end eta peaks are clearly separated \cite{kullander}.}{Performance of SEC}

\newpage ~
\thispagestyle{empty}

\emptydoublepage


\addcontentsline{toc}{section}{References}

\newpage ~
\thispagestyle{empty}
\emptydoublepage


\appendix
\section*{\hypertarget{appendix}{Appendix A}}
\rhead{APPENDIX A}
\lhead{APPENDIX A}
\label{appendix}
\addcontentsline{toc}{section}{Appendix A}
\renewcommand{\thetable}{\alph{table}}
\setcounter{table}{0}









\begin{center}
\begin{longtable}{|l|l|l|l|}
\caption{Results from test with $^{241}\mathrm{AmBe}$ source} \label{sourceRES} \\

\hline \multicolumn{1}{|c|}{\textbf{cable}} &
 \multicolumn{1}{c|}{\textbf{peak position [a.u.]}} &
 \multicolumn{1}{c|}{\textbf{peak sigma [a.u.]}} &
 \multicolumn{1}{c|}{\textbf{high voltage [V]}}\\ \hline 
\endfirsthead

\multicolumn{4}{c}%
{{\bfseries \tablename\ \thetable{} -- continued from previous page}} \\
\hline \multicolumn{1}{|c|}{\textbf{cable}} &
\multicolumn{1}{c|}{\textbf{peak position [a.u.]}} &
\multicolumn{1}{c|}{\textbf{peak sigma [a.u.]}} &
\multicolumn{1}{c|}{\textbf{high voltage [V]}}\\ \hline 
\endhead

\hline \multicolumn{4}{|r|}{{Continued on next page}} \\ \hline
\endfoot

\hline \hline
\endlastfoot
459 & 14.18 & 1.98 & 1615 \\
460 & 12.37 & 1.59 & 1521 \\
461 & 13.29 & 2.18 & 1900 \\
462 & 15.45 & 2.24 & 1575 \\
463 & 7.87 & 1.13 & 1440 \\
464 & 10.31 & 1.49 & 1550 \\
465 & 11.51 & 1.7 & 1640 \\
466 & 12.4 & 1.94 & 1639 \\
467 & 11.3 & 1.67 & 1510 \\
470 & 6.67 & 1.06 & 1526 \\
471 & 12.38 & 1.73 & 1720 \\
472 & 14.8 & 2.19 & 1620 \\
473 & 19.81 & 2.73 & 1639 \\
474 & 18.42 & 2.48 & 1613 \\
475 & 16.22 & 2.51 & 1703 \\
476 & 13.68 & 1.97 & 1622 \\
477 & 15.1 & 2.1 & 1620 \\
478 & 11.15 & 1.7 & 1572 \\
479 & 17.11 & 2.61 & 1719 \\
483 & 15.09 & 1.97 & 1480 \\
484 & 15.42 & 2.05 & 1583 \\
485 & 13.23 & 1.78 & 1465 \\
486 & 13.89 & 2.04 & 1535 \\
487 & 11.03 & 1.55 & 1587 \\
488 & 11.63 & 1.66 & 1566 \\
489 & 13.75 & 1.99 & 1650 \\
490 & 14.94 & 2.05 & 1480 \\
491 & 13.12 & 1.8 & 1610 \\
492 & 14.14 & 2.05 & 1640 \\
493 & 18.84 & 3.4 & 1750 \\
494 & 10.92 & 1.68 & 1644 \\
495 & 14.73 & 2.02 & 1750 \\
496 & 14.9 & 2.4 & 1680 \\
497 & 14.83 & 2.24 & 1695 \\
498 & 15.77 & 2.19 & 1638 \\
499 & 14.24 & 1.83 & 1466 \\
500 & 16.19 & 2.52 & 1678 \\
501 & 14.98 & 2.19 & 1592 \\
502 & 7.99 & 1.3 & 1616 \\
503 & 12.66 & 1.82 & 1633 \\
504 & 13.42 & 2.44 & 1587 \\
507 & 13.71 & 2.02 & 1518 \\
508 & 10.95 & 1.43 & 1538 \\
510 & 12.9 & 1.91 & 1785 \\
511 & 15.31 & 2.09 & 1599 \\
512 & 12.55 & 1.74 & 1415 \\
513 & 12.68 & 1.88 & 1575 \\
514 & 13.74 & 2 & 1451 \\
515 & 7.61 & 1.12 & 1481 \\
516 & 14.19 & 2.08 & 1640 \\
517 & 7.82 & 1.09 & 1427 \\
518 & 14.5 & 2.07 & 1560 \\
519 & 19.42 & 2.41 & 1763 \\
520 & 13.16 & 1.85 & 1631 \\
521 & 15.6 & 2.26 & 1665 \\
522 & 16.39 & 2.29 & 1531 \\
523 & 14.07 & 1.89 & 1465 \\
524 & 14.63 & 2.04 & 1560 \\
525 & 14.63 & 2.05 & 1540 \\
526 & 14.95 & 2.1 & 1670 \\
527 & 15.3 & 2.19 & 1480 \\
528 & 12.78 & 1.84 & 1692 \\
531 & 12.41 & 1.74 & 1665 \\
532 & 18.49 & 2.6 & 1577 \\
533 & 13.98 & 2.02 & 1560 \\
534 & 6.96 & 1.06 & 1499 \\
535 & 12.76 & 1.8 & 1485 \\
536 & 11.87 & 1.69 & 1614 \\
537 & 11.45 & 1.74 & 1739 \\
538 & 15.69 & 2.21 & 1559 \\
539 & 11.58 & 1.66 & 1512 \\
540 & 13.12 & 1.88 & 1618 \\
541 & 13.46 & 1.78 & 1462 \\
542 & 9.07 & 1.34 & 1531 \\
543 & 12.86 & 1.87 & 1500 \\
544 & 8.35 & 1.29 & 1468 \\
545 & 18.53 & 3.05 & 1755 \\
546 & 15.99 & 2.3 & 1688 \\
547 & 14.06 & 1.95 & 1624 \\
548 & 11.25 & 1.57 & 1387 \\
549 & 17.17 & 2.41 & 1764 \\
550 & 10.2 & 1.44 & 1474 \\
551 & 10.77 & 1.65 & 1626 \\
555 & 18.42 & 2.16 & 1620 \\
556 & 7.11 & 1.04 & 1550 \\
557 & 15.28 & 2.04 & 1625 \\
558 & 11.15 & 1.57 & 1492 \\
559 & 9.05 & 1.41 & 1638 \\
560 & 18 & 2.47 & 1630 \\
561 & 12.49 & 1.84 & 1500 \\
562 & 12.09 & 1.68 & 1546 \\
563 & 13.9 & 2.01 & 1598 \\
564 & 10 & 1.43 & 1620 \\
565 & 12.27 & 1.68 & 1485 \\
567 & 15.78 & 2.16 & 1603 \\
568 & 15.99 & 2.36 & 1520 \\
569 & 15.14 & 2.24 & 1661 \\
570 & 16.21 & 2.12 & 1618 \\
571 & 11.4 & 1.63 & 1566 \\
572 & 14.64 & 2.13 & 1690 \\
573 & 14.18 & 2.07 & 1625 \\
574 & 18.92 & 2.89 & 1793 \\
575 & 14.7 & 2.08 & 1390 \\
579 & 15.36 & 2.35 & 1930 \\
580 & 14.46 & 2.21 & 1705 \\
581 & 7.49 & 1.11 & 1482 \\
582 & 11.93 & 1.69 & 1880 \\
584 & 12.36 & 1.79 & 1518 \\
585 & 10.29 & 1.43 & 1395 \\
586 & 8.38 & 1.25 & 1604 \\
587 & 14.39 & 2.04 & 1638 \\
588 & 10.82 & 1.74 & 1582 \\
589 & 18.82 & 3.3 & 1722 \\
590 & 15.19 & 2.21 & 1599 \\
591 & 11.17 & 1.7 & 1884 \\
592 & 16.1 & 2.28 & 1670 \\
593 & 19.71 & 2.94 & 1743 \\
594 & 11.94 & 1.83 & 1733 \\
595 & 15.66 & 2.25 & 1706 \\
596 & 10.17 & 1.59 & 1621 \\
597 & 16.74 & 2.62 & 1813 \\
598 & 12.91 & 1.97 & 1894 \\
600 & 13.03 & 1.97 & 1672 \\
603 & 10.83 & 1.84 & 1521 \\
604 & 13.86 & 2.1 & 1725 \\
605 & 10.69 & 1.55 & 1593 \\
606 & 10.03 & 1.41 & 1510 \\
607 & 8.38 & 1.35 & 1477 \\
608 & 11.95 & 1.82 & 1491 \\
609 & 11.85 & 1.98 & 1743 \\
610 & 13.29 & 1.92 & 1583 \\
611 & 14.45 & 2.11 & 1691 \\
612 & 11.05 & 1.69 & 1566 \\
613 & 14.72 & 2.06 & 1688 \\
614 & 8.08 & 1.3 & 1519 \\
615 & 6.65 & 0.99 & 1717 \\
616 & 19.85 & 3.16 & 1600 \\
617 & 12.56 & 1.93 & 1698 \\
618 & 12.71 & 1.95 & 1680 \\
619 & 9.32 & 1.6 & 1693 \\
620 & 12.9 & 2.03 & 1739 \\
621 & 15.66 & 2.24 & 1812 \\
622 & 9.25 & 1.58 & 1821 \\
624 & 14.07 & 2.3 & 1762 \\
627 & 6.34 & 0.96 & 1561 \\
628 & 15 & 2.21 & 1655 \\
629 & 15.9 & 2.14 & 1725 \\
630 & 12.83 & 1.65 & 1682 \\
631 & 13.49 & 1.89 & 1544 \\
632 & 17.8 & 2.5 & 1740 \\
633 & 8.28 & 1.17 & 1481 \\
634 & 14.76 & 2.3 & 1700 \\
635 & 13.79 & 1.99 & 1636 \\
636 & 11.64 & 1.72 & 1615 \\
637 & 10.12 & 1.65 & 1771 \\
638 & 15.45 & 2.28 & 1640 \\
639 & 10.29 & 1.65 & 1578 \\
640 & 12.71 & 1.84 & 1697 \\
641 & 15.58 & 2.52 & 1756 \\
642 & 13.32 & 2.12 & 1697 \\
643 & 12.2 & 1.93 & 1603 \\
644 & 8.86 & 1.5 & 1676 \\
645 & 13.01 & 2.1 & 1600 \\
646 & 14.18 & 2.41 & 1843 \\
647 & 15.11 & 2.64 & 1882 \\
649 & 12.71 & 2.18 & 1745 \\
650 & 6.59 & 0.98 & 1484 \\
651 & 12.28 & 1.67 & 1826 \\
652 & 17.22 & 2.36 & 1828 \\
653 & 10.25 & 1.55 & 1560 \\
654 & 10.24 & 1.76 & 1655 \\
655 & 14.77 & 2.15 & 1697 \\
656 & 9.41 & 1.55 & 1515 \\
657 & 14.38 & 2.09 & 1656 \\
658 & 12.56 & 1.94 & 1479 \\
659 & 12.92 & 1.87 & 1774 \\
660 & 16.12 & 2.41 & 1672 \\
661 & 13.59 & 2.07 & 1550 \\
662 & 13.85 & 2.16 & 1582 \\
663 & 14.38 & 2.14 & 1582 \\
665 & 12.84 & 1.92 & 1727 \\
666 & 14.96 & 2.41 & 1631 \\
667 & 16.96 & 2.53 & 1600 \\
668 & 13.94 & 2.2 & 1645 \\
669 & 14.85 & 2.34 & 1755 \\
670 & 19.15 & 2.94 & 1847 \\
671 & 14.84 & 2.28 & 1580 \\
796 & 13.66 & 1.83 & 1570 \\
797 & 10.59 & 1.58 & 1600 \\
798 & 16.25 & 2.48 & 1270 \\
800 & 14.84 & 2.17 & 1388 \\
801 & 11.18 & 1.64 & 1758 \\
802 & 12.81 & 1.88 & 1737 \\
803 & 14.71 & 2.11 & 1713 \\
804 & 12.56 & 1.78 & 1561 \\
805 & 14.52 & 2.26 & 1761 \\
808 & 11.07 & 1.65 & 1638 \\
810 & 11.03 & 1.57 & 1777 \\
811 & 14.67 & 2.17 & 1817 \\
812 & 11.47 & 1.69 & 1602 \\
813 & 8.95 & 1.43 & 1624 \\
814 & 18.73 & 2.6 & 1675 \\
815 & 10.68 & 1.64 & 1631 \\
816 & 15.07 & 2.1 & 1558 \\
817 & 9.9 & 1.47 & 1500 \\
819 & 11.32 & 1.66 & 1815 \\
820 & 9.37 & 1.37 & 1469 \\
821 & 8.5 & 1.29 & 1474 \\
822 & 15.03 & 2.09 & 1772 \\
823 & 15.18 & 2.56 & 1664 \\
824 & 15.52 & 2 & 1599 \\
825 & 13.57 & 2.13 & 1372 \\
826 & 9.63 & 1.44 & 1502 \\
827 & 14.28 & 2.14 & 1548 \\
828 & 17.63 & 2.72 & 1669 \\
830 & 15.28 & 2.43 & 1648 \\
831 & 13.6 & 1.96 & 1539 \\
832 & 14.7 & 2.17 & 1427 \\
833 & 12.1 & 1.92 & 1618 \\
834 & 6.89 & 1.33 & 1420 \\
837 & 10.83 & 1.79 & 1746 \\
838 & 15.3 & 2.16 & 1795 \\
839 & 6.25 & 1.02 & 1552 \\
840 & 12.18 & 1.94 & 1610 \\
841 & 14.09 & 2 & 1600 \\
842 & 17.35 & 2.41 & 1579 \\
843 & 15.81 & 2.04 & 1547 \\
844 & 12.69 & 2.1 & 1758 \\
846 & 11.99 & 1.67 & 1654 \\
847 & 19.43 & 2.79 & 1685 \\
848 & 15.39 & 2.04 & 1427 \\
849 & 6.46 & 1.02 & 1665 \\
850 & 14.25 & 1.95 & 1549 \\
851 & 11.12 & 1.47 & 1540 \\
852 & 13.41 & 1.95 & 1686 \\
853 & 13.06 & 1.73 & 1567 \\
854 & 6.85 & 1.05 & 1436 \\
855 & 12.39 & 1.75 & 1536 \\
856 & 14.54 & 2.1 & 1583 \\
857 & 18.24 & 2.31 & 1540 \\
858 & 14.42 & 1.97 & 1709 \\
859 & 15.34 & 2.2 & 1626 \\
860 & 14.73 & 2.1 & 1611 \\
861 & 14.35 & 1.94 & 1770 \\
862 & 15.67 & 2.12 & 1740 \\
863 & 15.76 & 2.13 & 1640 \\
864 & 14.97 & 2.1 & 1702 \\
865 & 14.31 & 1.93 & 1475 \\
866 & 12.07 & 1.68 & 1548 \\
867 & 15.19 & 2.03 & 1628 \\
868 & 14.36 & 2.07 & 1622 \\
869 & 9.97 & 1.53 & 1503 \\
870 & 14.12 & 2.15 & 1572 \\
871 & 7.49 & 1.17 & 1571 \\
872 & 13.9 & 1.9 & 1522 \\
873 & 12.13 & 1.75 & 1599 \\
874 & 13.15 & 2.02 & 1652 \\
875 & 10.8 & 1.56 & 1572 \\
876 & 11.82 & 1.73 & 1557 \\
877 & 16.13 & 2.25 & 1752 \\
878 & 8.99 & 1.36 & 1426 \\
879 & 17.84 & 2.59 & 1667 \\
880 & 8.01 & 1.19 & 1526 \\
881 & 10.63 & 1.54 & 1528 \\
882 & 13.06 & 1.94 & 1502 \\
883 & 12.23 & 1.8 & 1532 \\
884 & 12.31 & 1.7 & 1554 \\
885 & 12.74 & 1.93 & 1492 \\
886 & 13.95 & 1.95 & 1618 \\
888 & 15.26 & 2.03 & 1657 \\
889 & 11.91 & 1.76 & 1454 \\
890 & 15.74 & 2.14 & 1627 \\
891 & 7.46 & 1.1 & 1406 \\
893 & 12.3 & 1.7 & 1717 \\
894 & 14.34 & 2.19 & 1610 \\
895 & 13.62 & 1.9 & 1619 \\
896 & 14.19 & 2.08 & 1644 \\
897 & 13.97 & 2.12 & 1692 \\
898 & 10.42 & 1.54 & 1570 \\
899 & 11.78 & 1.71 & 1622 \\
900 & 14.41 & 2 & 1608 \\
901 & 12.36 & 1.67 & 1585 \\
902 & 14.48 & 2.13 & 1516 \\
903 & 14.45 & 1.95 & 1658 \\
904 & 10.05 & 1.44 & 1394 \\
905 & 13.25 & 1.97 & 1645 \\
906 & 9.94 & 1.45 & 1562 \\
907 & 12.23 & 1.86 & 1674 \\
908 & 9.75 & 1.51 & 1464 \\
909 & 16.61 & 2.38 & 1490 \\
910 & 14.48 & 2.17 & 1755 \\
911 & 15.11 & 2.29 & 1680 \\
912 & 12.43 & 1.71 & 1549 \\
913 & 7.71 & 1.19 & 1528 \\
914 & 16.04 & 2.39 & 1637 \\
915 & 15.22 & 2.16 & 1650 \\
917 & 11.18 & 1.89 & 1740 \\
918 & 7.33 & 1.1 & 1682 \\
919 & 12.88 & 1.89 & 1600 \\
920 & 15.32 & 2.38 & 1560 \\
921 & 11.87 & 1.81 & 1610 \\
922 & 11.17 & 1.65 & 1664 \\
924 & 8.96 & 1.36 & 1625 \\
925 & 12.69 & 1.92 & 1813 \\
926 & 7.26 & 1.15 & 1418 \\
927 & 14.94 & 2.37 & 1712 \\
928 & 18.3 & 2.73 & 1800 \\
929 & 10.32 & 1.55 & 1669 \\
930 & 11 & 1.87 & 1749 \\
931 & 12.06 & 1.89 & 1750 \\
932 & 15.62 & 2.31 & 1627 \\
933 & 18.55 & 2.93 & 1752 \\
934 & 13.6 & 2 & 1814 \\
935 & 13.63 & 2.04 & 1766 \\
936 & 9.87 & 1.54 & 1821 \\
937 & 15.53 & 2.32 & 1558 \\
938 & 15.1 & 2.3 & 1676 \\
940 & 14.45 & 2.28 & 1729 \\
941 & 12.45 & 1.99 & 1763 \\
943 & 17.26 & 3.16 & 1820 \\
944 & 14.64 & 2.27 & 1713 \\
945 & 18.26 & 2.54 & 1711 \\
946 & 8.59 & 1.31 & 1586 \\
947 & 15.04 & 2.01 & 1646 \\
948 & 10.43 & 1.55 & 1684 \\
949 & 13.3 & 1.97 & 1411 \\
950 & 13.46 & 1.97 & 1740 \\
951 & 11.48 & 1.77 & 1632 \\
952 & 11.48 & 1.75 & 1730 \\
953 & 10.02 & 1.55 & 1623 \\
954 & 6.67 & 1.21 & 1761 \\
955 & 8.47 & 1.45 & 1794 \\
956 & 14.29 & 2.2 & 1676 \\
957 & 12.92 & 2.04 & 1712 \\
958 & 10.35 & 1.72 & 1765 \\
959 & 17 & 2.57 & 1780 \\
960 & 11.18 & 3.39 & 1705 \\
961 & 13.3 & 1.92 & 1626 \\
962 & 12.25 & 1.85 & 1838 \\
964 & 13.16 & 2.05 & 1782 \\
965 & 13.66 & 2.03 & 1656 \\
966 & 12.16 & 2.11 & 1777 \\
967 & 16.02 & 2.2 & 1780 \\
968 & 15.03 & 2.22 & 1709 \\
969 & 15.3 & 2.45 & 1615 \\
970 & 12 & 1.84 & 1653 \\
971 & 13.58 & 1.8 & 1678 \\
972 & 14.49 & 2.24 & 1769 \\
973 & 7 & 1.16 & 1591 \\
974 & 13.21 & 2.14 & 1691 \\
975 & 13.86 & 2.91 & 1664 \\
976 & 15.74 & 2.4 & 1821 \\
977 & 8.93 & 1.38 & 1565 \\
978 & 8.44 & 1.27 & 1662 \\
979 & 11.77 & 1.88 & 1656 \\
980 & 10.04 & 1.63 & 1750 \\
981 & 17.73 & 3.02 & 1447 \\
982 & 18.74 & 2.82 & 2034 \\
983 & 18.52 & 2.88 & 1994 \\
984 & 9.1 & 1.55 & 1648 \\
985 & 11.83 & 1.73 & 1698 \\
986 & 13.66 & 2 & 1653 \\
987 & 15.18 & 2.2 & 1600 \\
988 & 12.87 & 2.07 & 1698 \\
990 & 18.94 & 3.13 & 1819 \\
991 & 14.63 & 2.07 & 1700 \\
992 & 15.8 & 2.3 & 1534 \\
993 & 13.85 & 2 & 1621 \\
994 & 13.77 & 1.97 & 1568 \\
995 & 13.88 & 1.95 & 1616 \\
996 & 11.05 & 1.62 & 1450 \\
997 & 19.18 & 2.95 & 1718 \\
998 & 8.34 & 1.36 & 1523 \\
999 & 8.23 & 1.32 & 1492 \\
1000 & 11.92 & 1.91 & 1618 \\
1001 & 7.44 & 1.45 & 1481 \\
1002 & 13.8 & 2.84 & 1935 \\
1003 & 13.45 & 2.12 & 1985 \\
1005 & 15.9 & 2.56 & 1569 \\
1006 & 15.96 & 2.58 & 1892 \\
1007 & 13.69 & 2.2 & 1753 \\
1008 & 14.14 & 2.74 & 1920 \\
1009 & 14.16 & 2.94 & 1880 \\
1010 & 8.05 & 1.44 & 1497 \\
1011 & 6.31 & 1.27 & 1543 \\
1013 & 11.95 & 1.95 & 1695 \\
1014 & 15.06 & 2.56 & 1802 \\
1015 & 13.1 & 1.8 & 1800 \\
1016 & 8.05 & 1.52 & 1528 \\
1017 & 16.08 & 2.33 & 1685 \\
1018 & 14.74 & 2.16 & 1779 \\
1019 & 11.67 & 1.75 & 1638 \\
1020 & 13.87 & 2.41 & 1700 \\
1021 & 14.69 & 2.15 & 1675 \\
1022 & 16.61 & 2.33 & 1658 \\
1023 & 11.99 & 1.75 & 1772 \\
1024 & 13.36 & 2.03 & 1576 \\
1025 & 8 & 1.27 & 1492 \\
1026 & 13.91 & 2.37 & 1925 \\
1027 & 11.41 & 1.84 & 1668 \\
1028 & 13.51 & 2.34 & 1771 \\
1142 & 12.62 & 2.77 & 1550 \\
1143 & 12.5 & 1.88 & 1771 \\
1144 & 12.68 & 1.99 & 1484 \\
1145 & 10.46 & 1.56 & 1576 \\
1146 & 13.22 & 1.95 & 1709 \\
1147 & 13.87 & 2.11 & 1693 \\
1148 & 10.6 & 1.56 & 1541 \\
1149 & 15.86 & 2.62 & 1457 \\
1150 & 11.82 & 2.09 & 1724 \\
1151 & 8.71 & 1.49 & 1622 \\
1152 & 16.44 & 2.55 & 1794 \\
1153 & 13.42 & 1.94 & 1736 \\
1154 & 13.05 & 2.26 & 1603 \\
1155 & 12.85 & 2.04 & 1792 \\
1156 & 9.33 & 1.58 & 1635 \\
1157 & 13.05 & 2.15 & 1716 \\
1158 & 11.84 & 1.78 & 1629 \\
1159 & 15.71 & 2.43 & 1697 \\
1160 & 10.46 & 1.81 & 1826 \\
1161 & 13.14 & 2.24 & 1692 \\
1162 & 12.7 & 2 & 1723 \\
1164 & 12.86 & 2.07 & 1748 \\
1166 & 12.25 & 1.97 & 1526 \\
1167 & 9.32 & 1.48 & 1588 \\
1168 & 13.92 & 2 & 1610 \\
1169 & 10.75 & 1.73 & 1647 \\
1170 & 12.42 & 1.71 & 1611 \\
1172 & 11.51 & 1.67 & 1455 \\
1175 & 13.15 & 1.97 & 1656 \\
1176 & 12.39 & 1.96 & 1693 \\
1178 & 10.87 & 1.66 & 1544 \\
1179 & 16.33 & 2.47 & 1706 \\
1180 & 11.01 & 1.77 & 1609 \\
1181 & 12.43 & 2.07 & 1729 \\
1182 & 15.29 & 2.21 & 1698 \\
1183 & 13.88 & 2.01 & 1652 \\
1184 & 10 & 1.65 & 1702 \\
1186 & 10.87 & 1.73 & 1474 \\
1187 & 18.06 & 2.64 & 1704 \\
1188 & 8.54 & 1.32 & 1875 \\
1190 & 13.78 & 2 & 1762 \\
1192 & 14.01 & 2.11 & 1626 \\
1193 & 14.55 & 1.98 & 1617 \\
1194 & 9.48 & 1.42 & 1561 \\
1195 & 11.02 & 2.01 & 1577 \\
1196 & 11.3 & 1.6 & 1641 \\
1197 & 9.52 & 1.46 & 1677 \\
1198 & 11.81 & 1.84 & 1679 \\
1199 & 9.62 & 1.7 & 1666 \\
1200 & 12.06 & 1.81 & 1719 \\
1201 & 14.66 & 2.11 & 1712 \\
1202 & 10.72 & 1.52 & 1577 \\
1203 & 11.35 & 1.82 & 1801 \\
1204 & 18.05 & 2.7 & 1769 \\
1205 & 6.61 & 1.28 & 1749 \\
1206 & 11.86 & 1.93 & 1680 \\
1207 & 10.93 & 1.66 & 1550 \\
1208 & 14.18 & 2.07 & 1725 \\
1209 & 14.81 & 2.08 & 1560 \\
1210 & 9.21 & 1.43 & 1560 \\
1211 & 16.53 & 2.37 & 1621 \\
1212 & 10.56 & 1.73 & 1585 \\
1213 & 14.33 & 2.17 & 1569 \\
1214 & 15.19 & 2.28 & 1714 \\
1216 & 18.16 & 2.76 & 1681 \\
1217 & 8.21 & 1.25 & 1653 \\
1218 & 9.09 & 1.5 & 1648 \\
1219 & 6.75 & 1.1 & 1470 \\
1220 & 12.33 & 1.72 & 1712 \\
1221 & 18.14 & 2.88 & 1800 \\
1222 & 14.67 & 2.16 & 1789 \\
1223 & 13.43 & 2.02 & 1700 \\
1224 & 13.79 & 2.02 & 1550 \\
1225 & 6.79 & 1.01 & 1487 \\
1226 & 13.32 & 2.02 & 1551 \\
1227 & 9.76 & 1.45 & 1488 \\
1228 & 15.48 & 2.2 & 1630 \\
1229 & 11.24 & 1.8 & 1689 \\
1230 & 6.59 & 1.08 & 1474 \\
1231 & 15.35 & 2.21 & 1620 \\
1232 & 11.44 & 1.85 & 1650 \\
1233 & 12.43 & 1.94 & 1622 \\
1234 & 14.15 & 2.32 & 1587 \\
1235 & 14.46 & 2.16 & 1651 \\
1236 & 16 & 2.39 & 1513 \\
1237 & 9.29 & 1.5 & 1534 \\
1238 & 13.19 & 2.11 & 1686 \\
1240 & 15.81 & 2.04 & 1639 \\
1241 & 14.24 & 1.95 & 1580 \\
1242 & 9.61 & 1.37 & 1592 \\
1243 & 10.51 & 1.44 & 1699 \\
1244 & 12.61 & 1.8 & 1506 \\
1245 & 10.79 & 1.6 & 1603 \\
1246 & 11.34 & 1.6 & 1580 \\
1247 & 9.19 & 1.35 & 1580 \\
1248 & 10.27 & 1.52 & 1462 \\
1249 & 12.8 & 2.18 & 1668 \\
1250 & 9.23 & 1.34 & 1500 \\
1251 & 6.44 & 0.99 & 1600 \\
1252 & 12.15 & 1.67 & 1600 \\
1253 & 13.54 & 2.16 & 1640 \\
1254 & 10.91 & 1.79 & 1738 \\
1255 & 11.28 & 1.61 & 1609 \\
1256 & 10.66 & 1.63 & 1673 \\
1257 & 12.58 & 1.71 & 1515 \\
1258 & 16.28 & 2.31 & 1760 \\
1259 & 9.61 & 1.45 & 1500 \\
1260 & 7.88 & 1.3 & 1592 \\
1261 & 8.58 & 1.21 & 1624 \\
1262 & 14.48 & 2.33 & 1914 \\
1264 & 7.16 & 1.15 & 1533 \\
1265 & 11.84 & 1.78 & 1448 \\
1266 & 8.75 & 1.36 & 1561 \\
1267 & 11.85 & 1.74 & 1532 \\
1268 & 12.8 & 1.8 & 1647 \\
1269 & 10.03 & 1.36 & 1441 \\
1270 & 13.66 & 2.17 & 1780 \\
1271 & 12.95 & 1.81 & 1440 \\
1272 & 13.8 & 2.04 & 1564 \\
1273 & 18.82 & 2.63 & 1583 \\
1274 & 14.03 & 2.38 & 1609 \\
1276 & 13.6 & 1.85 & 1572 \\
1277 & 10.18 & 1.58 & 1634 \\
1278 & 8.02 & 1.12 & 1419 \\
1279 & 17.05 & 2.43 & 1609 \\
1280 & 13.84 & 2.05 & 1775 \\
1281 & 14.92 & 2.1 & 1432 \\
1282 & 9.46 & 1.42 & 1391 \\
1283 & 14.54 & 2.38 & 1800 \\
1284 & 14.38 & 2.07 & 1625 \\
1285 & 16.97 & 2.44 & 1733 \\
1286 & 13.33 & 2.05 & 1660 \\
1288 & 12.71 & 1.86 & 1729 \\
1289 & 11.88 & 1.77 & 1600 \\
1290 & 19.58 & 2.95 & 1601 \\
1291 & 6.75 & 1.12 & 1546 \\
1292 & 9.46 & 1.53 & 1704 \\
1293 & 13.05 & 1.92 & 1676 \\
1294 & 9.49 & 1.46 & 1421 \\
1295 & 12.45 & 1.8 & 1522 \\
1296 & 10.23 & 1.62 & 1554 \\
1297 & 11.74 & 1.63 & 1510 \\
1298 & 8.8 & 1.35 & 1538 \\
1299 & 12.3 & 1.81 & 1594 \\
1300 & 12.56 & 1.87 & 1427 \\
1301 & 12.03 & 1.71 & 1598 \\
1302 & 11.76 & 1.68 & 1504 \\
1303 & 13.55 & 1.9 & 1527 \\
1304 & 11.4 & 1.58 & 1513 \\
1305 & 9.49 & 1.41 & 1593 \\
1306 & 19.25 & 3.01 & 1605 \\
1307 & 12.5 & 1.75 & 1291 \\
1308 & 13.67 & 1.94 & 1549 \\
1309 & 14.55 & 2.17 & 1680 \\
1310 & 15.24 & 2.2 & 1542 \\
1312 & 6.81 & 1.13 & 1593 \\
1313 & 9.76 & 1.43 & 1500 \\
1314 & 16.05 & 2.36 & 1640 \\
1315 & 12.68 & 2.21 & 1779 \\
1316 & 11.1 & 1.74 & 1540 \\
1318 & 17.21 & 2.73 & 1630 \\
1319 & 12.55 & 2.92 & 2050 \\
1320 & 7.57 & 1.37 & 1704 \\
1321 & 16.89 & 2.73 & 1690 \\
1322 & 11.76 & 2.04 & 1502 \\
1323 & 18.23 & 2.58 & 1649 \\
1324 & 15.77 & 2.13 & 1631 \\
1325 & 11.9 & 2.09 & 1468 \\
1326 & 16.02 & 2.3 & 1675 \\
1327 & 8.68 & 1.51 & 1567 \\
1328 & 10.67 & 1.56 & 1565 \\
1329 & 15.35 & 2.35 & 1589 \\
1330 & 7.39 & 1.24 & 1559 \\
1331 & 14.73 & 2.25 & 1575 \\
1332 & 6.43 & 1.29 & 1600 \\
1333 & 13.79 & 2.07 & 1682 \\
1415 & 12.64 & 1.95 & 1600 \\
1416 & 12.04 & 1.87 & 1783 \\
1417 & 13.86 & 2.05 & 1610 \\
1418 & 13.35 & 1.99 & 1854 \\
1419 & 9.15 & 1.47 & 1698 \\
1420 & 12.38 & 2 & 1866 \\
1421 & 12.44 & 1.77 & 1667 \\
1422 & 15.74 & 2.82 & 1855 \\
1423 & 9.93 & 1.44 & 1564 \\
1424 & 13.4 & 2.21 & 1818 \\
1425 & 14.02 & 2.19 & 1649 \\
1426 & 18.07 & 2.59 & 1739 \\
1427 & 6.6 & 1.03 & 1543 \\
1428 & 17.36 & 2.47 & 1710 \\
1429 & 11.41 & 1.71 & 1581 \\
1430 & 9.72 & 1.48 & 1555 \\
1431 & 15.63 & 2.47 & 1780 \\
1432 & 11.84 & 2.17 & 1810 \\
1433 & 12 & 1.88 & 1773 \\
1434 & 9.94 & 1.63 & 1665 \\
1435 & 14.36 & 2.25 & 1564 \\
1436 & 9.38 & 1.45 & 1820 \\
1437 & 13.85 & 2.08 & 1611 \\
1438 & 15.18 & 2.21 & 1645 \\
1439 & 9.8 & 1.68 & 1631 \\
1440 & 15.26 & 2.35 & 1870 \\
1441 & 18.66 & 2.91 & 1770 \\
1442 & 8.48 & 1.25 & 1567 \\
1443 & 10.36 & 1.7 & 1606 \\
1444 & 10.97 & 1.9 & 1704 \\
1446 & 13.21 & 2.3 & 1855 \\
1447 & 12.65 & 1.91 & 1745 \\
1449 & 12.68 & 1.97 & 1654 \\
1450 & 14.49 & 2.09 & 1683 \\
1451 & 11.95 & 1.71 & 1544 \\
1452 & 14.03 & 2.21 & 1672 \\
1453 & 12.05 & 1.81 & 1784 \\
1454 & 16.23 & 2.36 & 1700 \\
1455 & 15.35 & 3.3 & 1837 \\
1456 & 8.82 & 1.42 & 1708 \\
1457 & 14.7 & 2.19 & 1778 \\
1458 & 13.19 & 2.02 & 1418 \\
1459 & 11.46 & 1.79 & 1633 \\
1460 & 10.75 & 1.71 & 1764 \\
1461 & 12.49 & 1.9 & 1797 \\
1462 & 6.27 & 1.19 & 1600 \\
1463 & 12.36 & 2.01 & 1585 \\
1464 & 11.9 & 2.08 & 1862 \\
1465 & 19.86 & 3.14 & 1750 \\
1466 & 12.46 & 2.17 & 1853 \\
1468 & 18.92 & 2.7 & 1800 \\
1470 & 14.45 & 2.46 & 1712 \\
1471 & 8.81 & 1.35 & 1490 \\
1472 & 14.66 & 2.17 & 1589 \\
1473 & 15.28 & 2.19 & 1630 \\
1474 & 10.97 & 2.14 & 1839 \\
1475 & 18.32 & 2.57 & 1671 \\
1476 & 13.54 & 2.03 & 1660 \\
1478 & 9.89 & 1.75 & 1842 \\
1479 & 13.23 & 1.91 & 1854 \\
1480 & 8.72 & 1.36 & 1613 \\
1481 & 18.25 & 2.64 & 1616 \\
1482 & 9.54 & 1.55 & 1714 \\
1483 & 7.32 & 1.25 & 1696 \\
1484 & 13.14 & 2.04 & 1681 \\
1485 & 15.37 & 2.51 & 1721 \\
1486 & 15.27 & 2.25 & 1600 \\
1487 & 11.59 & 2.26 & 1707 \\
1488 & 10.56 & 1.46 & 1676 \\
1489 & 19.83 & 2.97 & 1703 \\
1490 & 12.36 & 2.1 & 1627 \\
1491 & 17.86 & 2.6 & 1684 \\
1492 & 7.07 & 1.18 & 1428 \\
1493 & 10.12 & 1.54 & 1629 \\
1494 & 15.97 & 2.23 & 1582 \\
1495 & 12.08 & 1.73 & 1618 \\
1496 & 10.81 & 1.8 & 1694 \\
1497 & 7.87 & 1.22 & 1640 \\
1498 & 15.06 & 2.28 & 1583 \\
1499 & 13.05 & 1.83 & 1591 \\
1500 & 13.44 & 2.02 & 1663 \\
1501 & 7.98 & 1.12 & 1512 \\
1502 & 14.13 & 2.01 & 1714 \\
1503 & 14.07 & 1.99 & 1685 \\
1504 & 9.03 & 1.44 & 1605 \\
1505 & 12.62 & 1.64 & 1773 \\
1506 & 14.18 & 2.14 & 1550 \\
1507 & 13.41 & 1.9 & 1696 \\
1508 & 12.16 & 1.66 & 1660 \\
1509 & 13.77 & 2.03 & 1723 \\
1510 & 14.93 & 2.13 & 1548 \\
1511 & 13.3 & 1.94 & 1562 \\
1512 & 19.39 & 2.86 & 1800 \\
1513 & 15.63 & 2.32 & 1640 \\
1514 & 9.27 & 1.22 & 1497 \\
1515 & 10.84 & 1.59 & 1563 \\
1516 & 17.02 & 2.35 & 1713 \\
1517 & 12.42 & 1.95 & 1568 \\
1518 & 14.96 & 2.26 & 1712 \\
1519 & 10.51 & 1.43 & 1533 \\
1520 & 9.59 & 1.55 & 1612 \\
1521 & 8.59 & 1.29 & 1536 \\
1522 & 11.41 & 1.73 & 1611 \\
1523 & 7.92 & 1.23 & 1567 \\
1524 & 15.16 & 2.23 & 1578 \\
1525 & 15.41 & 2.24 & 1662 \\
1526 & 9.75 & 1.5 & 1405 \\
1527 & 10.55 & 1.86 & 1567 \\
1528 & 13.82 & 2.13 & 1489 \\
1529 & 9.78 & 1.46 & 1380 \\
1530 & 16.09 & 2.15 & 1644 \\
1531 & 10.76 & 1.64 & 1637 \\
1532 & 10.78 & 1.52 & 1551 \\
1532 & 12.17 & 1.7 & 1551 \\
1534 & 12.19 & 1.78 & 1495 \\
1535 & 11 & 1.72 & 1461 \\
1536 & 10.96 & 1.53 & 1379 \\
1537 & 17.46 & 2.33 & 1625 \\
1538 & 7.93 & 1.27 & 1339 \\
1539 & 12.19 & 1.92 & 1674 \\
1540 & 16.83 & 2.44 & 1648 \\
1541 & 6.15 & 0.93 & 1450 \\
1542 & 14.43 & 2.16 & 1658 \\
1543 & 12.05 & 1.77 & 1521 \\
1544 & 10.69 & 1.57 & 1371 \\
1545 & 10.49 & 1.55 & 1579 \\
1546 & 14.99 & 2.23 & 1726 \\
1547 & 13.63 & 2.05 & 1569 \\
1548 & 13.51 & 1.94 & 1515 \\
1549 & 16.46 & 2.24 & 1776 \\
1550 & 8.2 & 1.25 & 1405 \\
1551 & 12.97 & 1.89 & 1527 \\
1552 & 9.38 & 1.29 & 1491 \\
1553 & 13.18 & 2.01 & 1602 \\
1554 & 9.85 & 1.51 & 1514 \\
1555 & 9.79 & 1.53 & 1583 \\
1557 & 9.92 & 1.58 & 1452 \\
1558 & 12 & 1.72 & 1586 \\
1559 & 14.7 & 2.16 & 1492 \\
1560 & 14.61 & 2.1 & 1620 \\
1561 & 6.63 & 1.02 & 1526 \\
1562 & 11.64 & 1.67 & 1602 \\
1563 & 8.69 & 1.29 & 1601 \\
1564 & 14.56 & 2.21 & 1701 \\
1565 & 15.41 & 2.2 & 1700 \\
1566 & 16.04 & 2.33 & 1588 \\
1568 & 11.7 & 1.81 & 1651 \\
1569 & 7.31 & 1.14 & 1545 \\
1570 & 18 & 3.32 & 1900 \\
1571 & 18.16 & 2.65 & 1944 \\
1572 & 13.22 & 2.03 & 1604 \\
1573 & 19.78 & 3.36 & 1648 \\
1574 & 9.72 & 1.59 & 1816 \\
1575 & 13.15 & 1.98 & 1603 \\
1576 & 14.79 & 2.17 & 1696 \\
1577 & 12.01 & 1.9 & 1440 \\
1578 & 10.1 & 1.74 & 1386 \\
1579 & 16.43 & 2.51 & 1741 \\
1580 & 12.51 & 1.89 & 1426 \\
1581 & 12.88 & 2 & 1667 \\
1582 & 11.77 & 1.83 & 1699 \\
1583 & 9.77 & 1.61 & 1540 \\
1584 & 14 & 2.11 & 1661 \\
1585 & 12.94 & 1.96 & 1706 \\
1586 & 9.87 & 1.69 & 1473 \\
1587 & 13.7 & 2.09 & 1762 \\
1588 & 17.1 & 2.61 & 1822 \\
1589 & 14.69 & 2.47 & 1742 \\

\end{longtable}
\end{center}
\newpage ~
\thispagestyle{empty}
\emptydoublepage

\section*{Acknowledgements}
\thispagestyle{empty}
Here I would like to thank all people that helped me to create this thesis and without whom it wouldn't have been possible. 
I am not able to express my gratitude with my English, I will try my best.
\bigskip

First of all I would like to thank my supervisor Dr. hab. Zbigniew Rudy for introducing me the secrets of data analysis and for time spent on discussions, his guidance was always present.
\medskip

I am also gratefully to Dr. Volker Hejny, my tutor during the stay in J\"ulich, for showing me interesting aspects of calorimeter physics and for our fruitfull discussions and, of course, for new definition of common words. I would like
to thank Volker as well for being the indispensable source of information
concerning experimental physics. Thank You for patience!
\medskip

I would like also to thank Prof. Hans Str\"oher for possibility of stay in Forschungszentrum-J\"ulich and for giving me the opportunity to work with WASA~at~COSY collaboration.
\medskip

I am also very grateful to Prof. Reinhard Kulessa for allowing me to prepare this thesis in the Nuclear Physics Department of the Jagiellonian University.
\medskip

I want to express my appreciation to Prof. Lucjan Jarczyk, Prof. Bogusław Kamys and Prof. Ewa Gudowska-Nowak for their support of my stay in Forschungszentrum.
\medskip

I also thank all colleagues from WASA~at~COSY collaboration. It was a pleasure for me to work with You.
\medskip

I thank my colleagues in IKP: Alexey Dzyuba, Dieter Oellers for the joy of daily work.
\medskip

I thank also Pawel Podkopal and Michał Janusz for a great time outside the institute.
\medskip

I also want to express my gratitude to my mother Aniela and father Gerard and brother Michał(Leniwiec) and the rest of my Family, especially granny Rozalia and aunt Bożena for their great love, delicious meals and support through five years of studies in Cracow. Thank You!

\newpage ~
\thispagestyle{empty}
\emptydoublepage ~
\end{document}